\documentclass[journal]{./IEEEtran}

\ifCLASSOPTIONcompsoc
  \usepackage[nocompress]{cite}
\else
  \usepackage{cite}
\fi
\ifCLASSINFOpdf
\usepackage[pdftex]{graphicx}
\graphicspath{{./figs/}}
\else
\fi
\usepackage{amsmath}
\usepackage{amssymb}
\usepackage{algorithm}

\makeatletter
\renewcommand{\ALG@name}{}
\makeatother

\ifCLASSOPTIONcompsoc
 \usepackage[caption=false,font=footnotesize,labelfont=sf,textfont=sf]{subfig}
\else
 \usepackage[caption=false,font=footnotesize]{subfig}
\fi
\usepackage{url}
\usepackage{tcolorbox}
\tcbuselibrary{theorems}

\usepackage{relsize}

\DeclareMathSymbol{\shortminus}{\mathbin}{AMSa}{"39}
\DeclareMathOperator{\Ad}{Ad}
\DeclareMathOperator{\diag}{diag}
\DeclareMathOperator{\Cov}{\mathrm{Cov}}
\DeclareMathOperator{\R}{\mathbb{R}} 

\usepackage[acronym]{glossaries}
\newacronym{MC}{MC}{Motion Capture}
\newacronym{EKF}{EKF}{Extended Kalman Filter}
\newacronym{DLGEKF}{D-LG-EKF}{Discrete Lie-Group Extended Kalman Filter}
\newacronym{PDF}{PDF}{Probability Density Function}
\newacronym{PCB}{PCB}{Printed Circuit Board}
\newacronym{FIM}{FIM}{Fisher Information Matrix}
\newacronym{ML}{ML}{Maximum Likelihood}
\newacronym{MLE}{MLE}{Maximum Likelihood Estimator}
\newacronym{BCD}{BCD}{Block Coordinate Descent}
\newacronym{DOF}{DoF}{Degrees of Freedom}
\newacronym{MEMS}{MEMS}{Micro-Electro-Mechanical Systems}
\newacronym[plural=IMUs]{IMU}{IMU}{Inertial Measurement Unit}
\newacronym{RMSE}{RMSE}{Root-Mean-Square Error}
\newacronym{SNR}{SNR}{Signal-to-Noise Ratio}
\newacronym{GFIN}{GFIN}{Gyroscope-free Inertial Navigation}
\newacronym{GFIMU}{GFIMU}{Gyroscope-free Inertial Measurement Unit}
\newacronym{ODE}{ODE}{Ordinary Differential Equation}
\newacronym[plural=ZUPTs]{ZUPT}{ZUPT}{Zero-velocity Update}

\usepackage{tikz}
\usepackage{tikz-3dplot}
\usetikzlibrary{calc}
\usetikzlibrary{positioning}
\usepackage{pgfplots}
\usepackage{siunitx}

\usepackage{tikz-3dplot}

\usepackage[framemethod=tikz]{mdframed}
\usepackage{framed}

\usepackage{paralist}

\begin{document}
%
\title{Inertial Navigation Using an Inertial Sensor Array}

\author{H\aa kan~Carlsson,~\IEEEmembership{Student~Member,~IEEE,}
  Isaac~Skog,~\IEEEmembership{Senior~Member,~IEEE}
  Gustaf~Hendeby,~\IEEEmembership{Senior~Member,~IEEE}
  and~Joakim~Jald\'{e}n,~\IEEEmembership{Senior~Member,~IEEE}
\thanks{H.\,Carlsson is with the Department of Intelligent Systems, KTH, e-mail: hakcar@kth.se}%
\thanks{I.\,Skog is with the Department of Electrical
  Engineering, LiU, e-mail: isaac.skog@liu.se}%
\thanks{G.\,Hendeby is with the Department of Electrical
  Engineering, LiU, e-mail: gustaf.hendeby@liu.se}%
\thanks{J.\
  Jald\'{e}n is with the Department of Intelligent Systems, KTH,
  e-mail: jalden@kth.se}
\thanks{This work was partially supported by the Wallenberg Autonomous
    Systems and Software Program (WASP), and by the Swedish Foundation
    for Strategic Research (SSF) via the project ASSEMBLE.}
}



\IEEEtitleabstractindextext{%
\begin{abstract}
  We present a comprehensive framework for fusing
  measurements from multiple and generally placed accelerometers and
  gyroscopes to perform
  inertial navigation. Using the angular
  acceleration provided by the accelerometer array, we show
  that the numerical integration of the orientation can be done with second-order
  accuracy, which is more accurate compared
  to the traditional first-order accuracy that can be achieved when only
  using the gyroscopes. Since orientation errors are the
  most significant error source in inertial navigation,
  improving the orientation estimation reduces the overall
  navigation error. The practical performance benefit depends on prior
  knowledge of the inertial sensor array, and therefore we present four
  different state-space models using different underlying assumptions
  regarding the orientation modeling. The models are evaluated using a
  Lie Group Extended Kalman filter through simulations and
  real-world experiments.
  We also show how individual accelerometer biases are unobservable and
  can be replaced by a six-dimensional bias term whose dimension is
  fixed and independent of the number of accelerometers.
\end{abstract}



}

\maketitle

\IEEEdisplaynontitleabstractindextext

%
\IEEEpeerreviewmaketitle

\ifCLASSOPTIONcompsoc
\IEEEraisesectionheading{\section{Introduction}\label{sec:introduction}}
\else
\section{Introduction}
\label{sec:introduction}
\fi

\IEEEPARstart{I}{nertial} navigation is the process of estimating the
traveled distance and orientation of an object by time-integration of
measured velocities and accelerations~\cite{Titterton2004}. Due to the
integrative nature of inertial navigation, the estimated
position and orientation do inherently accumulate errors over
time. This unbounded accumulation of position and
orientation errors can only be limited by including external information
from other sensor systems that provide an absolute reference to the
environment or by including motion constraints. For instance, such aided
inertial navigation has been done using
satellite~\cite{Farrell2008}, video~\cite{Huang2019},
radio~\cite{Angelis2009}, LIDAR~\cite{Tang2015}, and magnetic
field data~\cite{Kok2013}. Otherwise, the position and
orientation errors' growth rate can be
reduced
by decreasing the measurement errors of the inertial sensors or using
additional motion information such as
\glspl{ZUPT}~\cite{Wahlstroem2021}. Improving the
sensor hardware~\cite{King1998} reduces
measurement errors, but this typically comes with increasing cost and
sensor size. Another approach to reducing the measurement error is to
fuse the measurements from a redundant amount
of inertial sensors to produce a virtual sensor with higher accuracy. In
particular, small \gls{MEMS}-based inertial sensors can be
cheaply fabricated~\cite{Shaeffer2013} and are therefore suited for
the construction of such sensor arrays~\cite{Nilsson2016,Givens2019}.



Fusing measurements from
multiple accelerometers dispersed on a rigid
body for inertial navigation has a long
history~\cite{Schuler1967,Padgaonkar1975,Chen1994a,Tan2001,Pachter2013}.
When no gyroscopes are employed, this system is usually referred to as an
\gls{GFIMU} or an accelerometer array. Since an accelerometer array provides
information on the angular velocity and the angular
acceleration, it can in principle replace gyroscopes for orientation
estimation. Omission of gyroscopes for inertial navigation was in
particular appealing for early \gls{MEMS}-based sensors, since early
\gls{MEMS}-gyroscopes had inadequate performance and high energy
consumption. However, the \gls{GFIMU} cannot
uniquely determine the instantaneous angular velocity and for low
dynamic rotational motion, the estimation error of the angular velocity is
high~\cite{Skog2016}. The angular acceleration also has to be
time-integrated to unambiguously estimate the angular
velocity~\cite{Wahlstroem2018b},
which imposes an extra time integration step compared to conventional
inertial navigation. The extra integration step increases the error
growth rate with an order of magnitude~\cite{Park2005} and for certain
configurations of the accelerometer
array the resulting non-linear \gls{ODE} has been reported to be
unstable~\cite{Padgaonkar1975,Nusbaum2017}. Mitigation of
these problems can be achieved by placing the accelerometers in
intricate configurations~\cite{Park2005,
  Hanson2005,Sahu2020,Williams2013,Chen1994a}. However, these
configurations may impose infeasible geometric constraints on the
system. Another approach has been to include a
gyroscope to the \gls{GFIMU}~\cite{Colomina2004, Williams2009, Bancroft2011,
  Patel2022, Waegli2008}, a type of
sensor assembly referred hereto as an inertial sensor array. To fuse
the measurements from such an inertial sensor array, an ML estimator was
presented in~\cite{Skog2016}. However, the
estimator assumed calibrated sensors and used an iterative method to
solve a non-linear optimization problem.

In this work, we present a comprehensive framework for fusing
measurements from multiple and generally placed accelerometers and
gyroscopes to perform
inertial navigation. The traditional
inertial navigation equations are extended to include the angular acceleration
provided by the
accelerometer array. The angular acceleration admits more accurate numerical
integration of the orientation with second-order accuracy compared to
the traditional
first-order accuracy that can be achieved when only using the
gyroscopes~\cite{Titterton2004}.
Since orientation errors are the
most significant error source in inertial navigation,
improving the orientation estimation will reduce the overall
navigation error. To practically benefit from the angular
acceleration, estimated by the accelerometer array, requires prior
knowledge of the inertial sensor
array. Therefore, we propose four state-space models, a.k.a
mechanization equations, using different underlying assumptions
regarding the time propagation of the orientation state. We evaluate
these models using a
Lie group Kalman filter through simulations and
experiments\footnote{Reproducible research: The data and the code used
  in the simulations and the experiments are available at
  \url{https://github.com/hcarlsso/Array-IN}}.
Using the
Kalman filter, we show that the non linear problem
of fusing the measurements
of multiple accelerometers and gyroscopes can be
solved by Kalman updates instead of using an iterative optimization
method~\cite{Skog2016}. Moreover, the Kalman updates are also shown
to stabilize the potentially unstable \gls{ODE}
when integrating the angular velocity. Finally, we note that
individual accelerometer biases
in an accelerometer array are unobservable~\cite{Waegli2008}. However,
we show that the individual accelerometer biases and the joint accelerometer
covariance matrix can be
replaced by a six-dimensional bias term and a six-by-six covariance
matrix. The dimensions of these new terms are fixed and
independent of the number of accelerometers.

\section{Inertial Navigation Modeling}
\label{sec:inert-navig-model}

This section presents the inertial navigation
equations. First, we introduce the continuous-time equations
for inertial sensor arrays to provide a rigorous derivation of
the second-order integration of the orientation enabled by the
array. Second, the continuous-time equations are discretized, and sensor models
for accelerometers and gyroscopes are presented.

\subsection{Continuous-time Inertial Navigation Equations}
\label{sec:cont-time-equat}

The inertial navigation equations describe a
moving body's time evolution relative to a frame at rest~\cite{Landau1976}.
This time-evolution or motion of the body is
computed from measurements given by inertial sensors attached to the
body. By convention, the
moving frame and the frame at rest are denoted as the body frame and the
navigation frame, respectively. Assuming that the body moves at such speeds that the Coriolis acceleration can
be neglected, then the continuous-time equations for inertial navigation
are given by~\cite{Titterton2004}
\begin{subequations}\label{eq:traditional_inertial_navigation}
\begin{align}
  \dot{R}^{nb} & = R^{nb}[\omega^b \times ], \label{eq:rotation_matrix_time_evolution_subscripts} \\
  \dot{p}^n & = v^n, \\
  \dot{v}^n & = g^n + R^{nb} s^b.
\end{align}
\end{subequations}
Here $p^n$ is the body's position in the navigation frame, $v^n$ is
the velocity, $g^n$ is the
local gravity\footnote{The centrifugal acceleration due to the earth's
rotation is assumed to be included in the local gravity
vector~\cite{Kok2017}.}, $\omega^b$ is the angular velocity between the navigation
frame and the body frame, and $s^b$ is the
specific force at the origin of the body
frame. The superscripts $n$ and $b$ denote the navigation and
body frame, respectively, and they indicate which coordinate frame a
quantity is expressed in. The rotation matrix between these two frames
is $R^{nb}$, which rotates a vector from the body
frame to the navigation
frame. Further, $[a \times ]b = a\times b$ is the skew-symmetric
matrix form of the cross-product. In conventional inertial navigation, the angular
velocity $\omega^b$ and
the specific force $s^b$ are measured by two sensor triads consisting
of three orthogonal gyroscope and accelerometer sensors,
respectively. Then~\eqref{eq:traditional_inertial_navigation} is
time-integrated to yield estimates of  $R^{nb}$, $p^n$, and $v^n$. The
location of the accelerometer triad defines the origin of the body frame.


Assuming the body to be rigid and have multiple accelerometer triads
geometrically dispersed on it, the specific force observed by
individual accelerometer triads will differ when the rigid body
undergoes a rotational
motion~\cite{Landau1976}. The specific force at a point not located in
the origin of the body frame, will have extra acceleration terms
due to the centrifugal and Euler accelerations. These extra
acceleration terms contain information about the rotation that can
extend~\eqref{eq:traditional_inertial_navigation}. Assuming that an
accelerometer triad is located
at $r_k^b$, the specific force at $r_k^b$ can then be related to the specific
force at the origin of the body frame $s^b$
via~\cite{Skog2016}
\begin{equation}
  \label{eq:pointwise_acceleration_triad}
  f_k^b = s^b + \underbrace{[\omega^b \times
    ]^2r_k^b}_{\substack{\text{Centrifugal}\\\text{acceleration}}} +
  \underbrace{[\dot{\omega}^b \times
    ]r_k^b}_{\substack{\text{Euler}\\\text{acceleration}}}.
\end{equation}
Here $\dot{\omega}^b$ is the angular acceleration of the rigid
body. Since $[\dot{\omega}^b \times ]r_k^b = -[r_k^b\times ]
\dot{\omega}^b $, it is noted that
\eqref{eq:pointwise_acceleration_triad} is linear in $s^b$ and
$\dot{\omega}^b$. Thus, by concatenating the measurements from
$K$ accelerometer triads using~\eqref{eq:pointwise_acceleration_triad} a
differential equation for $\dot{\omega}^b$ and $s^b$ can be obtained
as~\cite{Schuler1967,Park2005,Chen1994a,Tan2001,Pachter2013,Klein2015}
\begin{equation}\label{eq:array_inertial_navigation_general}
  \begin{bmatrix}
    \dot{\omega}^b \\ s^b
  \end{bmatrix}
  =
  \begin{bmatrix}
    A_{\dot{\omega}} \\ A_{s}
  \end{bmatrix}
  \left( f^b - h(\omega^b, r_{1:K}^{b}) \right),
\end{equation}
where $r_{1:K}^{b} \triangleq \{r_{k}^{b}\}_{k=1}^{K}$,
\begin{align}\label{eq:concat_specific_force}
  f^{b}
  &
\triangleq
\begin{bmatrix}
  f^{b}_1 \\
  \vdots  \\
  f^{b}_K
\end{bmatrix},
  &
    \text{and}
    &
  &
    h(\omega^b, r_{1:K}^{b})
  &
    \triangleq \begin{bmatrix}
    [\omega^b \times ]^2 r_1^b \\
    \vdots \\
    [\omega^b \times ]^2r_K^b
  \end{bmatrix}.
\end{align}
The matrices $A_{\dot{\omega}}$ and $A_{s}$ are functions of the
accelerometer positions $r_k^b$ and are further specified in
Appendix~\ref{sec:array-equations}. The interpretation
of~\eqref{eq:array_inertial_navigation_general} is that the difference
in the specific force at $r_k^b$ and the centrifugal acceleration
can be projected onto two different sub-spaces containing
$\dot{\omega}^b$ and $s^b$, respectively. If it can be
assumed that the accelerometer positions $r_k^b$ are centered in the
body frame, i.e., $\sum_{k=1}^{K} r_k^b =0$, then $A_{\dot{\omega}}$ and
$A_{s}$ become
\begin{subequations}\label{eq:def_A_general_centered}
\begin{align}
  A_{\dot{\omega}}
  &
    \triangleq
    \begin{bmatrix}
      A_1 & \cdots & A_K
    \end{bmatrix},
&
     A_{s}
  &
    \triangleq
    \frac{1}{K}
    \begin{bmatrix}
      I_3 & \cdots & I_3
    \end{bmatrix},
\end{align}
where
\begin{align}
  \label{eq:def_A_k}
  A_k & \triangleq  \left( \sum_{i=1}^K [ r_i^b \times ]^\top [ r_i^b
  \times ] \right)^{-1} [ r_k^b \times ],
\end{align}
\end{subequations}
as shown in Appendix~\ref{sec:array-equations}. Here $I_N$ is the
identity matrix of size $N$. For the inverse in
\eqref{eq:def_A_k} to be well defined, and in general for the matrix
$\begin{bmatrix}A_{\dot{\omega}}^\top & A_{s}^\top\end{bmatrix}^\top$
in \eqref{eq:array_inertial_navigation_general} to
have full rank, at least $K\geq 3$
accelerometer triads are required with positions that span a 2D
plane~\cite{Skog2016}. With $A_{\dot{\omega}}$ and
$A_{s}$ defined by \eqref{eq:def_A_general_centered} and
$\sum_{k=1}^{K} r_k^b =0$, it is also shown in
Appendix~\ref{sec:array-equations} that the differential
equations in~\eqref{eq:array_inertial_navigation_general} decouple
and become
\begin{subequations}\label{eq:diff_eq_continuous_array_ext}
  \begin{align}
    \dot{\omega}^b
    &
      =  \sum_{k = 1}^K A_k \left(  f_k^b - [\omega^b \times ]^2 r_k^b
      \right),\label{eq:angular_acceleration_body_frame}
    \\
    s^b & =  \frac{1}{K}  \sum_{k=1}^K  f_k^b  .    \label{eq:specific_force_body_frame}
  \end{align}
\end{subequations}
Hence, $s^b$ becomes independent of
$\omega^b$. Equation~\eqref{eq:specific_force_body_frame} could
equivalently been derived by taking the mean
of~\eqref{eq:pointwise_acceleration_triad} and assuming $\sum_{k=1}^{K} r_k^b =0$.

It has been reported that for certain arrangements of the accelerometer
positions and orientations the non linear \gls{ODE}
in \eqref{eq:angular_acceleration_body_frame} and
\eqref{eq:array_inertial_navigation_general} are
unstable\cite{Nusbaum2017, Schuler1967,Padgaonkar1975}. Since a gyroscope triad measures the angular velocity, a feedback law using
$\omega^b$ can be introduced to~\eqref{eq:angular_acceleration_body_frame} as
\begin{equation}\label{eq:angular_acceleration_body_frame_gyro_feedback}
  \dot{\omega}^b
  =  \sum_{k = 1}^K A_k \left(  f_k^b - [\omega^b \times ]^2 r_k^b
  \right) - L\omega^b,
\end{equation}
for a matrix $L$. The matrix $L$ can be designed as $L = \diag (l_x,
l_y,l_z)$, and for sufficiently high $l_i$, the poles in
\eqref{eq:angular_acceleration_body_frame_gyro_feedback} can be placed
in the left-hand plane~\cite{Khalil2002}. Thus, by providing feedback
from a gyroscope triad the potentially unstable \gls{ODE}
in~\eqref{eq:angular_acceleration_body_frame} can be stabilized.

In summary, using an accelerometer array with multiple accelerometers
triads dispersed on a rigid body, the inertial navigation equations in
\eqref{eq:traditional_inertial_navigation} can be extended with
\eqref{eq:diff_eq_continuous_array_ext}. Since the angular
acceleration can, via \eqref{eq:angular_acceleration_body_frame}, be
estimated only using the accelerometer triad measurements, it is
strictly speaking not necessary to include a gyroscope triad for
estimation of the rotation matrix in
\eqref{eq:rotation_matrix_time_evolution_subscripts}. However, by
including information from a gyroscope triad, the \gls{ODE} is ensured
to be stable.

\subsection{Discretized Inertial Navigation Equations}
\label{sec:line-discr}

The continuous-time equations need to be discretized before
implementation. When discretizing the traditional inertial
navigation equations in~\eqref{eq:traditional_inertial_navigation} it is commonly
assumed that the angular velocity and the specific
force are constant between each sample instant~\cite{Titterton2004}.
However, since the inertial sensor array also provides an
estimate of the angular acceleration, the discretization of the
rotation matrix in
\eqref{eq:rotation_matrix_time_evolution_subscripts} can instead
be based on the assumption of constant angular acceleration. For the
angular velocity, position, and velocity, which are expressed in
Euclidean vector-spaces, the discretization can be found by truncating
the Taylor series given by
\begin{subequations}
\begin{align}
  \omega(t + T)
  &
    = \omega(t) + \dot{\omega}(t)T + \mathcal{O}(T^2),
  \\
  p(t+T)
  &
    = p(t) + v(t)T + \dot{v}(t)\frac{T^2}{2} + \mathcal{O}(T^3),
  \\
    v(t+T)
    &
      = v(t) + \dot{v}(t)T + \mathcal{O}(T^2).
\end{align}
\end{subequations}
Here the superscripts related to the different frames have been
omitted for notational brevity. Further, $t$ and $T$ denote the time
and discretization (sample)
period, respectively. Moreover, $\mathcal{O}(T^n)$ denotes $n$:th and
higher-order terms of $T$. However, integration by Taylor
expansion is not possible for the rotation matrix $R$, since it
belongs to $SO(3) = \{ R \in \R^{3 \times 3}: R^\top R = I_3, \det R
= 1 \}$. That is, the orthogonality and determinant constraints
impose $SO(3)$ to be a 3-dimensional manifold embedded in $\R^{3
  \times 3}$. However, noting that $SO(3)$
is a matrix Lie group, Lie theory can be employed to integrate $R$ and
discretize
\eqref{eq:rotation_matrix_time_evolution_subscripts}. Instead of
integrating in $\R^{3 \times 3}$, the integration is performed in the
tangent space to $R$, also called the Lie algebra, which is isomorphic
to $\R^3$. For sufficiently small $T$ the solution
to~\eqref{eq:rotation_matrix_time_evolution_subscripts} is~\cite{Iserles2000}
\begin{equation}
  \label{eq:32}
  R(t+T) = R(t) \exp_{SO(3)}^{\land}(\theta(T) )
\end{equation}
where $\exp_{SO(3)}^{\land}$ is the matrix exponential map from $\R^3$
to $SO(3)$~\cite{Bourmaud2014}. Further, $\theta(T)$ is a vector
found as the solution to
\begin{equation}
  \label{eq:lie_algebra_diff_eq}
  \dot{\theta}(\tau) = \Gamma(\theta(\tau))\omega(t + \tau), \quad \theta(0) = 0,
\end{equation}
where
\begin{equation}
  \label{eq:27}
  \Gamma(\theta) = I_3 + \frac{1}{2}[\theta \times] + \left( 1 - \frac{\|\theta \|}{2} \cot\left(
        \frac{\| \theta \|}{2} \right) \right)\frac{1}{ \|\theta \|^2} [\theta \times]^2.
\end{equation}
In the navigation literature, $\Gamma(\theta(t))$ is also known as the
Bortz equation~\cite{Bortz1971, Shuster1993, Pittelkau2003}. The
physical interpretation of~\eqref{eq:32} is that
$\theta$ can be considered an orientation deviation or perturbation of
$R(t)$. The vector $\theta$ can also be interpreted as a rotation
vector that rotates $R(t)$. For $\tau \in [0,T]$, the
solution to~\eqref{eq:lie_algebra_diff_eq} is given by the Taylor series
\begin{equation}\label{eq:lie_algebra_diff_eq_solution}
  \theta(T) = \theta(0) + \dot{\theta}(0)T +
  \ddot{\theta}(0)\frac{T^2}{2} + \mathcal{O}(T^3).
\end{equation}
Thus the approximate solution to the integration problem in
\eqref{eq:lie_algebra_diff_eq} is given by calculating the
time-derivatives of $\theta(\tau)$ and $\Gamma(\theta(\tau))$. In
Appendix~\ref{sec:time-evol-rotat}, it is shown
that~\eqref{eq:lie_algebra_diff_eq_solution} is equal to
\begin{equation}
  \label{eq:9}
  \theta(T) = \omega(t)T + \dot{\omega}(t)\frac{T^2}{2} + \mathcal{O}(T^3).
\end{equation}
This result was also used in~\cite{Joukov2017} to integrate the
rotation matrix while including the second-order term, i.e.,
$\dot{\omega}$. Note again that in conventional inertial
navigation using an accelerometer and gyroscope triad the second-order term
would not be used due to the absence of direct measurements of
$\dot{\omega}$.

The discretized propagation equations for the navigation state are
obtained by omitting the higher-order terms in the Taylor series.
With a slight abuse of notation, we define the discrete samples as
$R_n  \triangleq R(t)$ and $R_{n+1} \triangleq R(t+T)$, and the other
variables similarly. The discretized equations for the navigation
state then become
\begin{subequations}\label{eq:discretized_navigation_equations}
\begin{align}
  R_{n+1}
    &
      = R_{n}\exp_{SO(3)}^{\land}\left( \omega_{n}T +
      \dot{\omega}_{n}\frac{T^2}{2}\right) ,\label{eq:rotation_discretized_2nd}
  \\
  \omega_{n+1}
  &
    = \omega_{n} + \dot{\omega}_{n}T, \label{eq:angular_velocity_discretized_2nd}
  \\
  p_{n+1}
  &
    = p_{n} + v_{n}T + (g+R_{n}s_n)\frac{T^2}{2}, \label{eq:position_discretized_2nd}
  \\
    v_{n+1}
    &
      = v_{n} + (g+R_{n}s_n)T. \label{eq:velocity_velocity_discretized_2nd}
\end{align}
\end{subequations}
Here, the navigation equations for the inertial sensor array
are propagated by the specific force $s_n$ and the angular
acceleration $\dot{\omega}_n$. Assuming that the angular velocity is
constant, that is $\dot{\omega}_n = 0$, and
removing \eqref{eq:angular_velocity_discretized_2nd},
then \eqref{eq:discretized_navigation_equations} reduces to the
conventional discretized inertial navigation
equations~\cite{Titterton2004}.

\subsection{Accelerometer Sensor Model}
\label{sec:sensor-modeling-accelerometers}

The accelerometer sensors are assumed to be calibrated in
terms of scale factors, misalignment, and cross-coupling~\cite{Carlsson2021}.
Further, the $k$:th accelerometer
triad's measurements at time-instant $n$ are assumed to be corrupted by
white noise and a slowly time-varying
bias~\cite{Xu2019a}. That is, the accelerometer measurements
$y_{k,n}^{(\text{a})}$ are modeled as
\begin{subequations}\label{eq:discrete_sensor_measurements_accelerometers}
  \begin{align}
    y_{k,n}^{(\text{a})}
    &
      = f_{k,n} + b_{k,n}^{(\text{a})} +
      w_{k,n}^{(\text{a})}  \label{eq:sensor_model_accelerometer},
    \\
    b_{k,n+1}^{(\text{a})}
    &
      = b_{k,n}^{(\text{a})} +
      w^{(b,\text{a})}_{k,n} , \label{eq:sensor_model_accelerometer_acc_bias}
\end{align}
\end{subequations}
Here $b_{k,n}^{(\text{a})}$, $w_{k,n}^{(\text{a})}$ and
$w^{(b,\text{a})}_{k}$ denote the
accelerometer triad's bias, measurement noise, and
driving
bias noise, respectively. The noise terms $w^{(\text{a})}_{k,n}$ and
$w^{(b,\text{a})}_{k,n}$ are assumed to be zero mean, uncorrelated in time,
uncorrelated with each others, and have
covariances $Q^{(\text{a})}_k$ and $Q^{(b,\text{a})}_k$, respectively.

Since the accelerometer triads' measurements only enter the
navigation equations through weighted sums in the terms for the
specific force $s_n$ and
the angular acceleration $\dot{\omega}_{n}$
in~\eqref{eq:diff_eq_continuous_array_ext}, the individual accelerometer
biases $b_{k,n}^{(\text{a})}$ will not be observable. To see
this, inserting~\eqref{eq:sensor_model_accelerometer} into
\eqref{eq:array_inertial_navigation_general} yields
\begin{equation}\label{eq:propation_specific_force_angular_acceleration_unreduced}
  \begin{bmatrix}
    \dot{\omega}_{n} \\
    s_n
  \end{bmatrix}
  =
  \begin{bmatrix}
    A_{\dot{\omega}}\\ A_{s}
  \end{bmatrix}
  \left(y_{n}^{(\text{a})} - b_{n}^{(\text{a})} - w_{n}^{(\text{a})} -
    h(\omega_n,r_{1:K}) \right)
\end{equation}
where
\begin{align}
  y_{n}^{(\text{a})}
  &
\triangleq
\begin{bmatrix}
  y_{1,n}^{(\text{a})} \\
  \vdots  \\
  y_{K,n}^{(\text{a})}
\end{bmatrix},
  &
    b_{n}^{(\text{a})}
  &
\triangleq
\begin{bmatrix}
  b_{1,n}^{(\text{a})} \\
  \vdots  \\
  b_{K,n}^{(\text{a})}
\end{bmatrix},
  &
    w_{n}^{(\text{a})}
  &
\triangleq
\begin{bmatrix}
  w_{1,n}^{(\text{a})} \\
  \vdots  \\
  w_{K,n}^{(\text{a})}
\end{bmatrix}.
\end{align}
From
\eqref{eq:propation_specific_force_angular_acceleration_unreduced}, it
is observed that $s_n$ and $\dot{\omega}_n$ depend on the
vectors
$b^{(\text{a})}$ and $w_{n}^{(\text{a})}$ only through the projection
matrix $\begin{bmatrix} A_{\dot{\omega}}^\top &
  A_{s}^\top \end{bmatrix}^\top \in \R^{6 \times 3K}$
defined in~\eqref{eq:def_A_general_centered}. The projection
matrix has full rank when $K\geq 3$ and when the accelerometer triads span a 2D
plane~\cite{Skog2016}. If the full rank assumption is fulfilled, the vectors
$b^{(\text{a})}\in \R^{3K} $ and $w_{n}^{(\text{a})} \in \R^{3K}$ are
projected to the
lower dimensional subspace $\R^{6}$, meaning that there exists $3K-6$
dimensional subspaces of $b^{(\text{a})}$ and $w_{n}^{(\text{a})}$
that are unobservable. The dimension of the state-vector
can thus be reduced to decrease the computational complexity and this
also solves the problem of unobservable bias terms for individual
accelerometers~\cite{Waegli2008}. Similarly, the noise vector
$w_n^{(a)}$ can be reduced. Define the new bias terms $b^{(\dot{\omega})}$ and
$b^{(s)}$ and the new noise terms $w^{(\dot{\omega})}$ and $w^{(s)}$ as
\begin{equation}
  \label{eq:bias_transformation}
  \begin{bmatrix}
    b^{(\dot{\omega})}_n \\ b^{(s)}_n
  \end{bmatrix}
  \triangleq
  -
  \begin{bmatrix}
    A_{\dot{\omega}}\\ A_{s}
  \end{bmatrix}
  b^{(\text{a})}_n,
  \quad
  \begin{bmatrix}
    w^{(\dot{\omega})}_n \\ w^{(s)}_n
  \end{bmatrix}
  \triangleq
  -
  \begin{bmatrix}
    A_{\dot{\omega}}\\ A_{s}
  \end{bmatrix}
  w^{(\text{a})}_n.
\end{equation}
Then
\eqref{eq:propation_specific_force_angular_acceleration_unreduced} can
be rewritten as
\begin{equation}\label{eq:propation_specific_force_angular_acceleration_reduced}
  \begin{bmatrix}
    \dot{\omega}_{n} \\
    s_n
  \end{bmatrix}
  =
  \begin{bmatrix}
    A_{\dot{\omega}}\\ A_{s}
  \end{bmatrix}
  \left(y_{n}^{(\text{a})} - h(\omega_n,r_{1:K})
    \right) +
  \begin{bmatrix}
    b_{n}^{(\dot{\omega})} \\ b_{n}^{(s)}
  \end{bmatrix} + \begin{bmatrix}
    w_{n}^{(\dot{\omega})} \\ w_{n}^{(s)}
  \end{bmatrix},
\end{equation}
where the function $h$ is given
in~\eqref{eq:array_inertial_navigation_general}. The propagation
equation for the new bias terms are
\begin{subequations}\label{eq:reduced_acc_bias}
\begin{align}
  b^{(\dot{\omega})}_{n+1}
  &
    = b^{(\dot{\omega})}_{n} + w^{(b,\dot{\omega})}_{n}, \\
  b^{(s)}_{n+1} & = b^{(s)}_{n} +  w^{(b,s)}_{n},
\end{align}
\end{subequations}
where the new driving bias noise terms $w^{(b,\dot{\omega})}_{n}$
and $w^{(b,s)}_{n}$ are introduced. The transformation
in~\eqref{eq:bias_transformation} will cause a correlation between
$w^{(\dot{\omega})}_{n}$ and $w^{(s)}_{n}$. The joint covariance of
for $w^{(\dot{\omega})}_n$ and $w^{(s)}_w$ is
\begin{equation}
  \label{eq:covariance_reduce_white_noise}
  \Cov\left( \begin{bmatrix}
      w^{(\dot{\omega})} \\ w^{(s)}
  \end{bmatrix} \right)
=
\begin{bmatrix}
  A_{\dot{\omega}}Q^{(\text{a})}A_{\dot{\omega}}^\top
  &
  A_{\dot{\omega}}Q^{(\text{a})}A_{s}^\top
  \\
  A_{s}Q^{(\text{a})}A_{\dot{\omega}}^\top
  &
  A_{s}Q^{(\text{a})}A_{s}^\top
  \end{bmatrix}.
\end{equation}
The joint covariance for the new terms $w^{(b,\dot{\omega})}_n$ and
$w^{(b,s)}_w$ can be derived
analogously. From~\eqref{eq:propation_specific_force_angular_acceleration_reduced},
we
note first that the dimension of the new bias and noise terms
are $6$ and independent of $K$. Second, it can
be noted that the covariance matrix $Q^{(a)}$ that is used to
propagate~\eqref{eq:array_inertial_navigation_general} is
replaced with the covariance matrix
in~\eqref{eq:covariance_reduce_white_noise}. The former has size $3K
\times 3K$ and the latter has size $6\times 6$, meaning that the same
noise structure can be represented with fewer elements.

Moreover, given the assumption of centered accelerometer positions,
i.e., $\sum r_k =0$, and assuming that $Q^{(\text{a})} =
\sigma_{a}^2 I_{3K}$, then~\eqref{eq:covariance_reduce_white_noise}
simplifies to
\begin{equation}
  \label{eq:covariance_acc_reduced}
  \Cov\left( \begin{bmatrix}
      w^{(\dot{\omega})} \\ w^{(s)}
  \end{bmatrix} \right)
=
\sigma_{a}^2
\begin{bmatrix}
  \left( \sum_{k=1}^K [ r_k \times ]^\top [ r_k \times] \right)^{-1}
  &
  0_{3,3}
  \\
  0_{3,3}
  &
  \frac{1}{K}I_3
\end{bmatrix}.
\end{equation}
In this case, the cross-correlation terms
vanish. Further, it can be observed that the covariance of $w^{(s)}$ is
inversely proportional to the number of accelerometer triads in the
array and independent of the geometry. This is the same result
as obtained in~\cite{Skog2016}, and the covariance for the specific force is the
same as the Cram\'{e}r-Rao lower bound. Moreover, assuming that the
accelerometer triads are mounted in a planar square grid with
spacing $\alpha$, the covariance for $w^{(\dot{\omega})}$ is inversely
proportional to $\alpha^2$, also a conclusion reached in~\cite{Skog2016}.

\subsection{Gyroscope Sensor Model}
\label{sec:gyrosc-sens-model}

The gyroscopes sensors are, similarly to the accelerometer sensors,
assumed to be assembled in triads, and measure the angular velocity in
three orthogonal directions. Moreover, the gyroscopes are also assumed to
be calibrated in terms of scale factors, misalignment, and
cross-coupling~\cite{Carlsson2021}.
Since the angular velocity is identical for all points on a rigid
body~\cite{Landau1976}, the measurements from multiple geometrically
dispersed gyroscope triads can be fused to create an equivalent
virtual gyroscope
triad~\cite{Wahlstroem2016}.
Hence, for the purpose of this work, it can without loss of generality
be assumed that there is only one
gyroscope triad in the inertial sensor array. The
measurements $y_{n}^{(\text{g})}$ from
the equivalent virtual gyroscope triad can be modeled as
\begin{subequations}\label{eq:discrete_sensor_measurements_gyroscopes}
  \begin{align}
    y_{n}^{(\text{g})}
    &
      = \omega_{n} + b_{n}^{(\text{g})} + w_{n}^{(\text{g})},\\
    b_{n+1}^{(\text{g})}
    &
      = b_{n}^{(\text{g})} + w^{(b,\text{g})}_{n} .
\end{align}
\end{subequations}
Here $b_{}^{(\text{g})}$, $w_{}^{(\text{g})}$ and $w^{(b,\text{g})}$ are the
gyroscope triad's bias, measurement noise, and driving
bias noise, respectively. The noise terms
$w^{(\text{g})}_{n}$  and $w^{(b,\text{g})}$,
are assumed to be zero mean, uncorrelated in time, and have
covariances $Q^{(\text{g})}$ and $Q^{(b,\text{g})}$, respectively.

\section{Inertial Sensor Array Navigation Equations}
\label{sec:model-variations}

This section presents the discretized inertial navigation equations
and the sensor models combined into different state-space models. The
models are constructed based on specific
assumptions on the prior knowledge of the inertial sensor array. This
section is concluded with a discussion on the advantages and
disadvantages of the different models.

\subsection{State-Space Models}
\label{sec:state-space-models-1}

\begin{figure*}
  \begin{mdframed}
  \begin{minipage}[t]{0.49\columnwidth}
    \input{alg_2nd_array_2}
    \vspace{10pt}
  \end{minipage}
  \hfill
  \begin{minipage}[t]{0.49\columnwidth}
    \input{alg_1st_array_2}
  \end{minipage}
  \begin{minipage}[t]{0.49\columnwidth}
    \input{alg_2nd_gyr_2}
  \end{minipage}
  \hfill
  \begin{minipage}[t]{0.49\columnwidth}
    \input{alg_1st_gyr_2}
  \end{minipage}
\end{mdframed}
\caption{Equations for the four proposed state-space models for
  inertial navigation using an inertial sensor array. The 2nd order
  models have a 2nd order term in the propagation of the rotation
  matrix in
  \eqref{eq:2nd_accelerometer_array_rotation} and
  \eqref{eq:2nd_gyroscope_rotation}, while the
  1st order models in \eqref{eq:1st_accelerometer_array_rotation} and
  \eqref{eq:1st_gyroscope_rotation} have a 1st order term. The accelerometer
  array models propagate the angular velocity in
  \eqref{eq:2nd_accelerometer_array_angular_velocity} and
  \eqref{eq:1st_accelerometer_array_angular_velocity} using the
  angular acceleration computed from the accelerometer measurements
  in \eqref{eq:2nd_accelerometer_array_omega_dot} and
  \eqref{eq:1st_accelerometer_array_omega_dot}. The gyroscope models
  estimate
  the angular velocity directly using \eqref{eq:2nd_gyroscope_angular_velocity}
  and \eqref{eq:1st_gyroscope_angular_velocity}.}\label{fig:state_space_models}
\end{figure*}

In the inertial sensor array, both the accelerometer array
measurements and the gyroscope measurements can be used to estimate
the angular velocity. The
accelerometer array achieves this through time-integration of the
angular acceleration,
and the gyroscopes measure it directly. Since orientation
estimation is an essential part of the inertial navigation process that amounts
to time-integration of the angular velocity, we can
either choose to use the accelerometer array, the gyroscope, or
fuse both to estimate the orientation. Here, we
present four different discrete state-space models shown in
Fig.~\ref{fig:state_space_models} of how to perform
inertial navigation with an accelerometer array and a gyroscope triad, using
different underlying assumptions regarding the time propagation of the
orientation state.

In the first state-space model
in~\eqref{eq:2nd_order_accelerometer_array}, we incorporate
all available information from both the
accelerometer array and the gyroscope triad by collecting
\eqref{eq:discretized_navigation_equations},
\eqref{eq:propation_specific_force_angular_acceleration_reduced},
\eqref{eq:reduced_acc_bias}, and
\eqref{eq:discrete_sensor_measurements_gyroscopes}. In this
model, the accelerometer array measurements are used to compute
the angular acceleration, which propagates both the angular velocity
and the rotation matrix using a 1st order and a 2nd order term,
respectively. This is analogous to the specific force propagating the
velocity and the
position in \eqref{eq:velocity_velocity_discretized_2nd} and
\eqref{eq:position_discretized_2nd}. Thanks to the 2nd
order term, the model is 2nd order accurate in the numerical integration of
the rotation matrix. This model
in~\eqref{eq:2nd_order_accelerometer_array} is from
heron referred to as the \textit{2nd order accelerometer array}
model.

If the accelerometer array is not sufficiently calibrated in terms of
accelerometer positions, scale factors, misalignment, or cross-coupling,
the estimation error of the angular acceleration might be
prohibitively high. There are two ways of reducing the impact of
a high estimation error in the angular acceleration in this
case. First, the rotation
matrix could be propagated only using a first-order term, which could
be more accurate even though the numerical integration is now only 1st
order accurate. This model is denoted the
\textit{1st order accelerometer array} model and is shown
in~\eqref{eq:1st_order_accelerometer_array}.
The second option is to omit
the propagation of the angular velocity while keeping the 2nd
order term for the rotation matrix. This is possible since the angular
acceleration in~\eqref{eq:angular_acceleration_body_frame} can be
computed with the angular velocity given by the gyroscope measurements
instead.
The integration of the rotation matrix is thus 2nd order accurate,
and this model is denoted as the \textit{2nd order gyroscope} model and
is shown in~\eqref{eq:2nd_order_gyroscope}. Here it is noted
that the angular acceleration bias $b^{(\dot{\omega})}$ is not propagated,
since that
would include two bias terms for the
propagation of $R_n$
in~\eqref{eq:2nd_gyroscope_rotation}, and would thus not be
observable. But we could have also omitted the gyroscope bias
$b^{(\text{g})}_{n}$ and kept $b^{(\dot{\omega})}_n$.

Moreover, the standard inertial navigation assumption could also be used, that
is, the rotation matrix is propagated only using a 1st order term
which is given by the gyroscopes. This model is
denoted as the \textit{1st order gyroscope} model and is shown
in~\eqref{eq:1st_order_gyroscope}.
The only difference between this model and the standard inertial
navigation equations is that the
specific force is now calculated using the mean value of the
accelerometer measurements.

\subsection{Discussion on State-Space Models}
\label{sec:disc-state-space-1}

The main difference between the accelerometer array and gyroscope
state-space models is whether the angular velocity is propagated or
not in the navigation equations. In the
accelerometer array models, the accelerometer array measurements are the sole
input to the inertial navigation process and used both for calculating
the translational and rotational changes. Hence,
\eqref{eq:2nd_accelerometer_array_rotation} to
\eqref{eq:2nd_accelerometer_array_specific_force} and
\eqref{eq:1st_accelerometer_array_rotation} to
\eqref{eq:1st_accelerometer_array_specific_force} can be
used to realize a \gls{GFIMU} inertial navigation
system~\cite{Chen1994a,Klein2015,Nusbaum2017,Pachter2013}. The
accelerometer array models could thus be used for high dynamic
applications where gyroscopes usually saturate~\cite{Pamadi2004,Camarillo2013}. If
rotational information from gyroscopes are to be included this must be
done via some filtering framework, such as the Kalman filter. In that
case, \eqref{eq:2nd_accelerometer_array_rotation} to
\eqref{eq:2nd_accelerometer_array_specific_force} and
\eqref{eq:1st_accelerometer_array_rotation} to
\eqref{eq:1st_accelerometer_array_specific_force}, describe the system
dynamics and \eqref{eq:2nd_accelerometer_array_gyro_measurement} and
\eqref{eq:1st_accelerometer_array_gyro_measurement} the
observation equation of the state-space model used in the filtering
process. A natural consequence of using the Kalman filter is that
accelerometer and gyroscope measurements
are fused using Kalman updates that have a fixed amount of
computations. This is not the case
for the current state-of-the-art method that fuses the
accelerometer and gyroscope measurements by solving a non linear optimization
problem using an iterative method~\cite{Skog2016}. The Kalman updates will also
function as a feedback mechanism on $\omega$, which has the added
benefit of stabilizing
the potentially
unstable continuous-time equation
in~\eqref{eq:angular_acceleration_body_frame_gyro_feedback}~\cite{Nusbaum2017,Padgaonkar1975}.
Moreover, it is also
possible to fuse the gyroscope and
accelerometer array measurements for orientation estimation without
Kalman updates as shown
in the 2nd order gyroscope model. However, the gyroscope models will
not work when the gyroscopes saturate.

\section{Kalman Filter for Array Inertial Navigation}
\label{sec:kalm-filt-inert}

An inertial navigation solution is usually used in conjunction with
other sensor systems. In this case, the sensor measurements are fused
in a filtering framework, which is the topic of this section. Since the
rotation matrix belongs to $SO(3)$, we present a
Lie group Kalman filter for inertial navigation using the state-space models
presented in Fig.~\ref{fig:state_space_models}. The filter
equations for a general matrix Lie group are
first presented, and then the specific
equations for the 2nd order accelerometer array state-space model in
\eqref{eq:2nd_order_accelerometer_array} are
presented. From this model, the other state-space models follow naturally.

\subsection{Discrete Lie Group Extended Kalman Filter}
\label{sec:lie-group-extended}

Since the rotation matrix belongs to $SO(3)$ the
standard Kalman filter cannot be directly used with the state-space
models in Fig.~\ref{fig:state_space_models}. Here we instead
employ the~\gls{DLGEKF}~\cite{Bourmaud2013} that explicitly
considers the non-Euclidean geometry of $SO(3)$. Since $\R^N$
can be embedded into a matrix, the
Euclidean space is also a matrix Lie group~\cite{Bourmaud2014}. And
since the composition of matrix Lie groups also is a matrix Lie group,
we only present the \gls{DLGEKF} for a single matrix Lie group
$G$. The \gls{DLGEKF} reduces to
the standard \gls{EKF} when $G$ is the
Euclidean space~\cite{Bourmaud2013}.

The \gls{DLGEKF} framework is based on the concept of concentrated
Gaussian distributions on Lie groups~\cite{Wang2006}. The random variable $X$
defined on the Lie group $G$ has a
Gaussian distribution with almost all of its probability mass concentrated in a small
neighborhood around the mean $\hat{X}$, so that the probability
distribution is effectively contained in the Lie
algebra to $\hat{X}$. Since the Lie algebra is isomorphic to
$\R^{N_x}$, the Gaussian distribution can be represented by the Euclidean random
variable $e \sim \mathcal{N}(0_{N_x,1}, P)$ where $P \in \R^{N_x \times N_x}$ is a
covariance matrix. Here $0_{N,M}$ is the zero matrix with $N$ rows and
$M$ columns.
The concentrated Gaussian
distribution on $G$ is then defined as
\begin{equation}
  \label{eq:concentrated_Gaussian}
  X = \hat{X} \exp_{G}^{\land}\left( e \right).
\end{equation}
Next, it is assumed in the  \gls{DLGEKF} framework that the
state-space model is in the form
\begin{subequations}\label{eq:LG_state_space}
\begin{align}
  X_{n+1}
  &
    = X_{n} \exp_{G^{}}^{\land}\left( \Omega(X_{n}, u_n, w_{n}^{(\text{p})})  \right), \\
  Y_n
  &
    = \eta(X_n) \exp_{G'}^{\land}\left( w_{n}^{(\text{m})} \right),
\end{align}
\end{subequations}
where $w_{n}^{(\text{p})} \in \R^{N_p}$ is the process
noise, $w_{n}^{(\text{m})} \in \R^{N_m}$ is the
measurement noise, and $u_n \in \R^{N_u}$ is the input.
The propagation function is $\Omega: G \times \R^{N_u} \times \R^{N_p}
\rightarrow \R^{N_x} $ and the measurement function is $\eta: G \rightarrow
G'$, as given by, e.g.,
\eqref{eq:2nd_order_accelerometer_array}. Next, it is assumed that the state
estimate of $X_{n}$ to be a concentrated Gaussian with mean
$\hat{X}_{n|n}$ and covariance $P_{n|n}$ according to
\eqref{eq:concentrated_Gaussian}, and that $w_{n}^{(\text{p})} \sim \mathcal{N}(0,
Q^{(\text{p})})$ and $w_{n}^{(\text{m})} \sim
\mathcal{N}(0,Q^{(\text{m})})$. Then, the propagation equations for the mean
and the covariance are
\begin{subequations}\label{eq:D-LG-EKG-propagation}
\begin{align}
  \hat{X}_{n+1|n}
  &
    = \hat{X}_{n|n} \exp_G^{\land}\left( \hat{\Omega}_{n} \right)
  \\
  P_{n+1|n}
  &
    = F_{n} P_{n|n} F_{n}^\top + G_nQ^{(\text{p})}G_n^\top
\end{align}
\end{subequations}
where
\begin{subequations}\label{eq:D-LG-EKG-propagation-jacobians}
\begin{align}
  \hat{\Omega}_{n}
  &
    =    \Omega(\hat{X}_{n|n}, u_n, 0), \\
  F_{n}
  &
    = \Ad_G\left( (\exp_G^{\land}\left( \shortminus\hat{\Omega}_{n} \right) \right) +
    \Phi_{n}(\hat{\Omega}_{n})J_n^{(x)},
  \\
    G_n
  &
    = \Phi_{G}(\hat{\Omega}_{n}) J_n^{(w)},
  \\
  J_n^{(x)}
  &
    = \left. \frac{\partial}{\partial e} \Omega\left(\hat{X}_{n|n}
    \exp_G^{\land}\left( e \right), u_n, 0 \right)
    \right\rvert_{e =
    0}, \label{eq:D-LG-EKG-propagation-jacobians-state}
  \\
  J_n^{(w)}
  & = \left. \frac{\partial}{\partial w} \Omega\left(\hat{X}_{n|n} ,
    u_n, w \right)
    \right\rvert_{w = 0}. \label{eq:D-LG-EKG-propagation-jacobians-noise}
\end{align}
\end{subequations}
Here $\Ad_G$ is the adjoint representation of $G$ on $\R^{N_x}$ and
$\Phi_{G}$ is the right Jacobian of $G$~\cite{Bourmaud2014}. The
measurement update is
\begin{subequations}\label{eq:D-LG-EKG-update}
\begin{align}
  \hat{X}_{n|n}
  &
    = \hat{X}_{n|n-1}\exp_G^{\land}\left( e_{n|n} \right)
  \\
  P_{n|n}
  &
    = \Phi_{G}( e_{n|n}) \left( I - K_n H_n  \right)
    P_{n|n-1} \Phi_{G}( e_{n|n})^\top
\end{align}
\end{subequations}
where
\begin{subequations}\label{eq:D-LG-EKG-update-jacobians}
\begin{align}
  K_n
  &
    = P_{n|n-1} H_n^{\top} \left( H_n P_{n|n-1} H_n^{\top}  + Q^{(\text{m})}
    \right)^{-1} \label{eq:D-LG-EKG-update-kalman-gain}
  \\
  e_{n|n}
  &
    = K_n \log_{G'}^{\lor}\left( \eta(\hat{X}_{n|n-1})^{-1} Y_n  \right)
  \\
  H_n
  & =
  \shortminus \left. \frac{\partial}{\partial e} \log_{G'}^{\lor}\left(
    \eta \left( \hat{X}_{n|n-1} \exp_{G}^{\land} \left( e \right)
    \right)^{-1} Y_n \right) \right\rvert_{e = 0}.
\end{align}
\end{subequations}
Here $\log_{G'}^{\lor}$ is the matrix logarithm of $G'$.
If $G$ and $G'$ were Euclidean spaces,
\eqref{eq:D-LG-EKG-propagation} to
\eqref{eq:D-LG-EKG-update-jacobians} would be
reduced to the regular
\gls{EKF} filter equations~\cite{Bourmaud2013, Bourmaud2014}. Compared to the
standard \gls{EKF} filter equations, we can note the
inclusion of the factors $\Phi_{G}(\hat{\Omega}_{n})$ and
$\Phi_{G}(e_{n|n})$ for the computation of the covariance
matrices. If $G = SO(3)$ these factors account for the rotation of the body
frame and the correction of the covariance
matrix for the rotation is also referred to as attitude reset~\cite{Pittelkau2003}.


\subsection{D-LG-EKF for Array Inertial Navigation}
\label{sec:kalman-filter-array}

Here we define the specific Lie-Group $G$ applicable for the
inertial navigation equations. We derive the filter for the 2nd order
accelerometer array state-space model in
\eqref{eq:2nd_order_accelerometer_array} and omit the other
state-space models, since they can be derived from the 2nd order accelerometer
array state-space model through deleting rows and columns in the
Jacobians. For the state-space model in
\eqref{eq:2nd_order_accelerometer_array} the matrix Lie group is
defined as ${G = SO(3) \times \R^{18}}$. The explicit matrix expression
for $G$ is found from the matrix embedding of the Euclidean elements
by letting
\begin{equation}
  \label{eq:14}
  z_n \triangleq
\begin{bmatrix}
  \omega_n^\top & p_n^\top & v_n^\top & (b^{(\dot{\omega})}_{n})^\top &
    (b^{(s)}_{n})^\top & (b_{n}^{(\text{g})})^\top
\end{bmatrix}^{\top}
\end{equation}
and
\begin{align}\label{eq:6}
  X_n
  &
    \triangleq
    \begin{bmatrix}
      R_n  & 0_{3,18} & 0 \\
      0_{18,3}  & I_{18} & z_n \\
      0_{1,3} &  0_{1,18} & 1
  \end{bmatrix}
  =
  \begin{bmatrix}
    R_n \\ \omega_n \\ p_n \\ v_n \\ b^{(\dot{\omega})}_{n} \\
    b^{(s)}_{n} \\ b_{n}^{(\text{g})}
  \end{bmatrix}_G.
\end{align}
The dimension of the Lie
algebra of $G$ is consequently 21 and the state-covariance $P \in
\R^{21 \times 21}$.  Moreover, the process noise vector is composed of
the accelerometer measurement noise and the white noise that drives
the bias terms for the accelerometers and gyroscopes, that is,
\begin{equation}
  \label{eq:15}
  w_n^{(\text{p})}  \triangleq
  \begin{bmatrix}
    w_{n}^{(\dot{\omega})} \\
    w_{n}^{(s)} \\
    w_{n}^{(b,\dot{\omega})} \\
    w_{n}^{(b, s)} \\
    w_{n}^{(b,\text{g})}
  \end{bmatrix}.
\end{equation}
The input to the propagation equation is the accelerometer
measurements. 
The propagation equation is then derived
from~\eqref{eq:2nd_order_accelerometer_array} as
\begin{equation}
  \label{eq:20}
  \Omega(X_n, y_{n}^{(\text{a})}, w_n^{(\text{p})}) =
  \begin{bmatrix}
    \omega_{n}T + \dot{\omega}_{n}\frac{T^2}{2}  \\
    \dot{\omega}_{n}T \\
    v_{n}T + (g + R_ns_n)\frac{T^2}{2} \\
    (g + R_ns_n)T \\
    w_{n}^{(b,\dot{\omega})} \\
    w_{n}^{(b,s)} \\
    w_{n}^{(b,\text{g})}
  \end{bmatrix},
\end{equation}
where $\dot{\omega}_{n}$ and $s_n$ are computed using
\eqref{eq:2nd_accelerometer_array_omega_dot} and
\eqref{eq:2nd_accelerometer_array_specific_force}.
The Jacobians $F_n$ and $G_n$ are given in
Appendix~\ref{sec:jacobians}.


Moreover, the angular velocity is propagated using the
accelerometer measurements in
\eqref{eq:2nd_accelerometer_array_angular_velocity}, but also provided
by the gyroscopes in
\eqref{eq:2nd_accelerometer_array_gyro_measurement}. Hence, the
gyroscope measurements have to be fused using Kalman updates
in~\eqref{eq:D-LG-EKG-update}. The Lie-Group for gyroscope
measurements is defined as $G^{(\text{g})} \triangleq \R^3$ and the
gyroscope measurement equation and its corresponding Jacobian is
\begin{subequations}\label{eq:D-LG-EKF-updates-gyro}
  \begin{align}
    \begin{bmatrix}
      I_3  & y_n^{(\text{g})} \\
      0_{1,3}  & 1  \\
    \end{bmatrix}
           &
             =
             \underbrace{\begin{bmatrix}
               I_3  & \omega_n  +  b^{(\text{g})}_{n} \\
               0_{1,3}  & 1  \\
             \end{bmatrix}}_{\triangleq \eta^{(\text{g})}(X_n)}
    \begin{bmatrix}
               I_3  & w_n^{(\text{g})}\\
               0_{1,3}  & 1  \\
             \end{bmatrix}
    \\
    H^{(\text{g})} & =
    \begin{bmatrix}
      0_{3} &  I_3 & 0_{3, 12} & I_{3}
    \end{bmatrix}.
  \end{align}
\end{subequations}
Here $0_{N}$ is the zero matrix of size $N$. As mentioned in
Section~\ref{sec:disc-state-space-1},
the non linear optimization problem in~\cite{Skog2016} is now solved using
Kalman updates in \eqref{eq:D-LG-EKG-update} and \eqref{eq:D-LG-EKF-updates-gyro}.
The benefit of using Kalman updates is that the required number of
computations are known in advance, which is not the case for iterative
optimization methods.

Furthermore, a measurement update on the position can be made if the inertial
navigation system has information on the position, e.g., when
using a GPS sensor. The measurement equation and the corresponding Jacobian
is given as
\begin{subequations}
\begin{align}
  \label{eq:position_updates}
  \begin{bmatrix}
      I_3  & y_n^{(p)} \\
      0_{1,3}  & 1  \\
    \end{bmatrix}
           &
             =
             \underbrace{\begin{bmatrix}
               I_3  & p_n   \\
               0_{1,3}  & 1  \\
             \end{bmatrix}}_{\triangleq \eta^{(p)}(X_n)}
    \begin{bmatrix}
               I_3  & w_n^{(p)}\\
               0_{1,3}  & 1  \\
             \end{bmatrix}
    \\
  H^{(p)}  & =  \begin{bmatrix}
    0_{3, 6} & I_3 & 0_{3, 12}
  \end{bmatrix}.
\end{align}
\end{subequations}
Here $y_n^{(p)}$ is the position measurement and $w_n^{(p)}$ is
position measurement noise.


\section{Simulations and Experiments}
\label{sec:simul-exper}

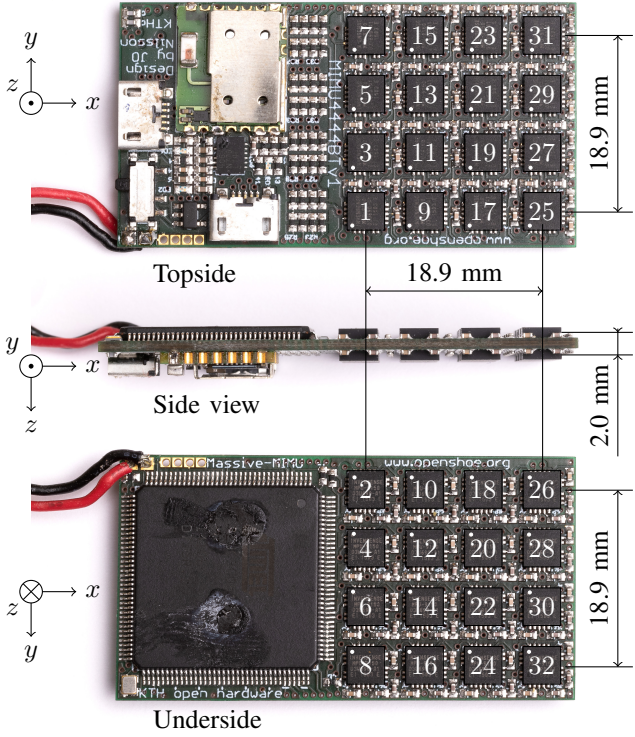
\begin{figure}
\begin{center}
\begin{tikzpicture}
\node[anchor=south west, inner sep=0] (image) at (0,0){\includegraphics[width=8.3cm]{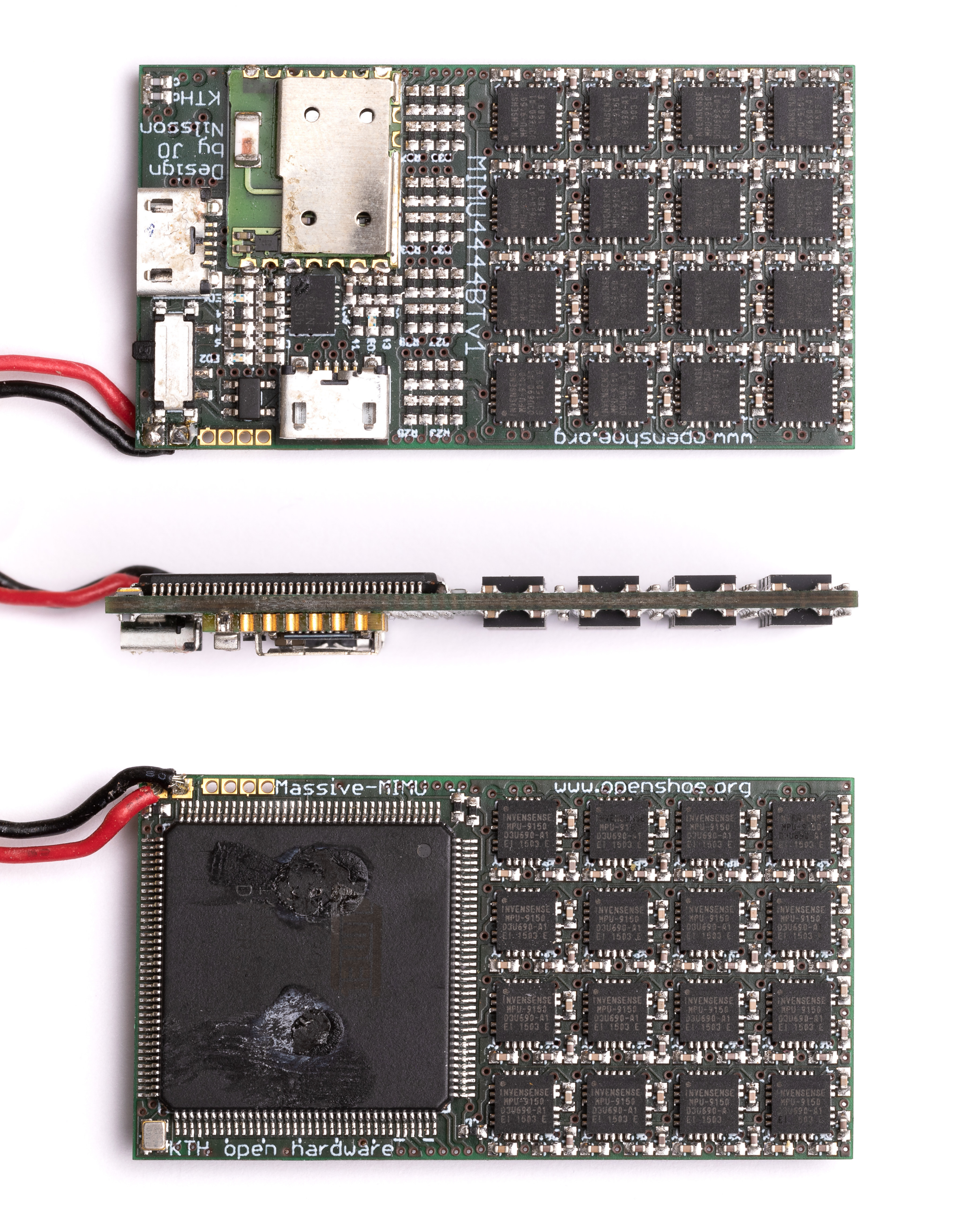}};
\begin{scope}[xshift=0.0cm,yshift=8.5cm]
\draw [->] (0,0) -- +(0,0.6) node [above] {$y$};
\draw [->] (0,0) -- +(0.6,0) node [right] {$x$};
\node [draw,shape=circle,inner sep=0,minimum size=3mm, fill=white] at (0,0) {};
\node [draw,shape=circle,inner sep=0,minimum size=0.5mm, fill=black] at (0,0) {};
\node [anchor=south east, inner sep = 1.5mm] at (0,0) {$z$};
\end{scope}
\begin{scope}[xshift=0.0cm,yshift=5.0cm]
\draw [->] (0,0) -- +(0,-0.6) node [below] {$z$};
\draw [->] (0,0) -- +(0.6,0) node [right] {$x$};
\node [draw,shape=circle,inner sep=0,minimum size=3mm, fill=white] at (0,0) {};
\node [draw,shape=circle,inner sep=0,minimum size=0.5mm, fill=black] at (0,0) {};
\node [anchor=south east, inner sep = 1.5mm] at (0,0) {$y$};
\end{scope}
\begin{scope}[xshift=0.0cm,yshift=2.0cm]
\draw [->] (0,0) -- +(0,-0.6) node [below] {$y$};
\draw [->] (0,0) -- +(0.6,0) node [right] {$x$};
\node [draw,shape=circle,inner sep=0,minimum size=3mm, fill=white] at (0,0) {};
\draw (1.0607mm,1.0607mm) -- (-1.0607mm,-1.0607mm);
\draw (1.0607mm,-1.0607mm) -- (-1.0607mm,1.0607mm);
\node [anchor=north east, inner sep = 1.5mm] at (0,0) {$z$};
\end{scope}
\node [anchor=west] at (1.5,6.2) {Topside};
\node [anchor=west] at (1.5,4.5) {Side view};
\node [anchor=west] at (1.5,0.3) {Underside};
\draw (4.45,6.8) -- (4.45,3.6);
\draw (6.8,6.89) -- (6.80,3.6);
\draw [<->] (4.45,6.0) -- (6.80,6.0) node [midway,above] {18.9 mm};
\draw (7.1,7.05) -- (8.0,7.05);
\draw (7.1,9.40) -- (8.0,9.40);
\draw [<->] (7.8,7.05) -- (7.8,9.40) node [midway,above,sloped] {18.9 mm};
\draw (7.05,5.45) -- (8.0,5.45);
\draw (7.05,5.15) -- (8.0,5.15);
\draw [->] (7.8,5.5) -- (7.8,5.45);
\draw [-] (7.8,5.45) -- (7.8,5.15);
\draw [->] (7.8,3.8) -- (7.8,5.15) node [midway,above,sloped] {2.0 mm};
\draw (7.1,1.00) -- (8.0,1.00);
\draw (7.1,3.35) -- (8.0,3.35);
\draw [<->] (7.8,1.00) -- (7.8,3.35) node [midway,above,sloped] {18.9 mm};

\begin{scope}[xshift=4.45cm,yshift=1.02cm]
  \foreach \x in {0,1,2,3} {
    \foreach \y in {0,1,2,3} {
      \node at ($(\x*7.8mm,\y*7.8mm)$) [rectangle,draw=black,minimum
      size=5.0mm,inner sep=0mm,fill=none] {\color{white}
        \pgfmathparse{2*(4*\x  +  3-\y + 1) }%
        \pgfmathprintnumber{\pgfmathresult}%
      };
    }
  }
\end{scope}

\begin{scope}[xshift=4.45cm,yshift=70.6mm]
  \foreach \x in {0,1,2,3} {
    \foreach \y in {0,1,2,3} {
      \node at ($(\x*7.8mm,\y*7.8mm)$) [rectangle,draw=black,minimum
      size=5.0mm,inner sep=0mm,fill=none] {\color{white}
        \pgfmathparse{2*(4*\x  +  \y + 1) - 1 }%
        \pgfmathprintnumber{\pgfmathresult}%
      };
    }
  }
\end{scope}

\draw[white,fill=white] (4.27,6.87) circle (0.2mm);
\draw[white,fill=white] (4.28,7.64) circle (0.2mm);
\draw[white,fill=white] (4.29,8.42) circle (0.2mm);
\draw[white,fill=white] (4.30,9.21) circle (0.2mm);
\draw[white,fill=white] (5.03,6.87) circle (0.2mm);
\draw[white,fill=white] (5.04,7.64) circle (0.2mm);
\draw[white,fill=white] (5.05,8.42) circle (0.2mm);
\draw[white,fill=white] (5.06,9.21) circle (0.2mm);
\draw[white,fill=white] (5.83,6.87) circle (0.2mm);
\draw[white,fill=white] (5.84,7.64) circle (0.2mm);
\draw[white,fill=white] (5.85,8.42) circle (0.2mm);
\draw[white,fill=white] (5.86,9.21) circle (0.2mm);
\draw[white,fill=white] (6.63,6.87) circle (0.2mm);
\draw[white,fill=white] (6.64,7.64) circle (0.2mm);
\draw[white,fill=white] (6.65,8.42) circle (0.2mm);
\draw[white,fill=white] (6.66,9.21) circle (0.2mm);
\draw[white,fill=white] (4.27,1.20) circle (0.2mm);
\draw[white,fill=white] (4.28,2.00) circle (0.2mm);
\draw[white,fill=white] (4.29,2.77) circle (0.2mm);
\draw[white,fill=white] (4.30,3.52) circle (0.2mm);
\draw[white,fill=white] (5.03,1.20) circle (0.2mm);
\draw[white,fill=white] (5.04,2.00) circle (0.2mm);
\draw[white,fill=white] (5.06,2.76) circle (0.2mm);
\draw[white,fill=white] (5.06,3.52) circle (0.2mm);
\draw[white,fill=white] (5.83,1.20) circle (0.2mm);
\draw[white,fill=white] (5.83,1.99) circle (0.2mm);
\draw[white,fill=white] (5.85,2.76) circle (0.2mm);
\draw[white,fill=white] (5.86,3.52) circle (0.2mm);
\draw[white,fill=white] (6.62,1.20) circle (0.2mm);
\draw[white,fill=white] (6.63,1.99) circle (0.2mm);
\draw[white,fill=white] (6.62,2.76) circle (0.2mm);
\draw[white,fill=white] (6.63,3.52) circle (0.2mm);
\end{tikzpicture}
\end{center}
\caption{Layout and dimensions of the inertial sensor array used
  in the experiments and simulations, along with the numbering of the
  \gls{IMU} triads used in tables and figures. The dimensions denote
  the distance between the center positions of the \gls{IMU} chip packages. The white dots
  on each \gls{IMU} package are the orientation markers,
  indicating the mounting directions of the \gls{IMU} packages~\cite{2013-MPU-9150-spec}. The
  \gls{IMU} packages on
  the underside, even numbers, are rotated \num{180}\si{\degree} around the axis $y =
  x$ relative to the \gls{IMU} packages on the topside, odd numbers. Each
  \gls{IMU} package measures $\num{4}~\si{mm} \times \num{4}~\si{mm} \times \num{1}~\si{mm}$~\cite{2013-MPU-9150-spec}.}
\label{fig:array}
\end{figure}

The performance of the proposed Kalman filter and the four proposed
state-space models for inertial navigation
using an inertial sensor array was evaluated by
simulations and experiments. The
simulations demonstrate the performance obtained under idealized
assumptions, and real-world experiments
verify the practical applicability of the proposed methods.

\subsection{Simulations}
\label{sec:simulations}

\begin{figure}[!t]
  \centering
  \subfloat[High dynamics\label{fig:sim_w_motion_high}]{%
    \includegraphics[width=\columnwidth]{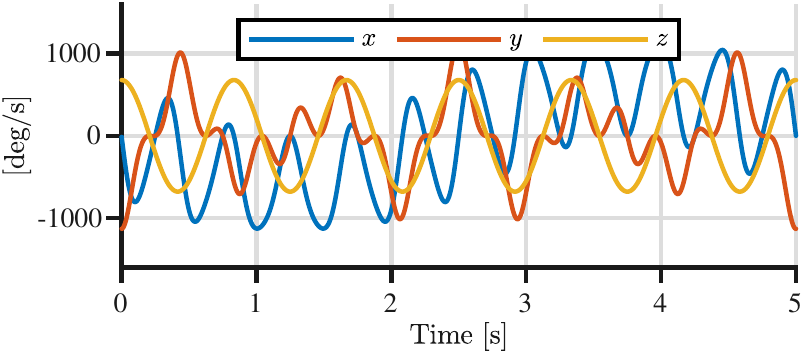}
  }
  \\
  \subfloat[Low dynamics\label{fig:sim_w_motion_low}]{%
    \includegraphics[width=\columnwidth]{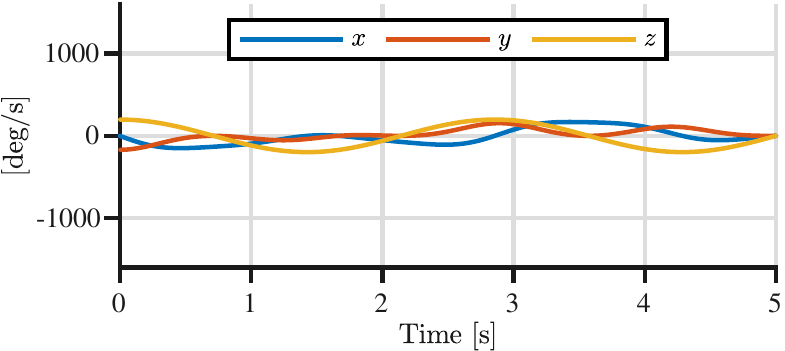}
  }
  \caption{The angular velocity for the motion used in the
    simulations, where \protect\subref{fig:sim_w_motion_high} and
    \protect\subref{fig:sim_w_motion_low} represent high and low rotational
    dynamics, respectively.}\label{fig:simulations_angular_velocity}
\end{figure}

\begin{figure*}[!t]
  \centering
    \subfloat[Low dynamics, $F_s = 500$ Hz\label{fig:simulations_pos_rmse_low_dynamics_high_Fs}]{%
    \includegraphics[width=\columnwidth]{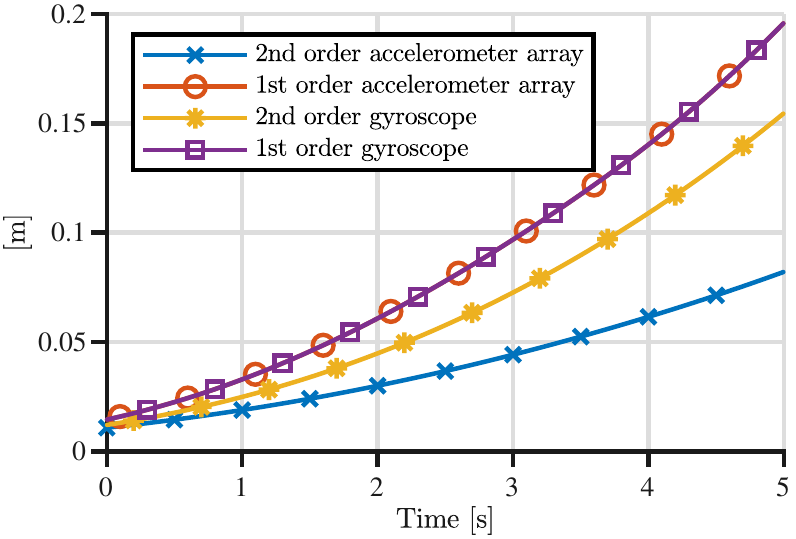}
  }
  \hfil
    \subfloat[High dynamics, $F_s = 500$ Hz\label{fig:simulations_pos_rmse_high_dynamics_high_Fs}]{%
    \includegraphics[width=\columnwidth]{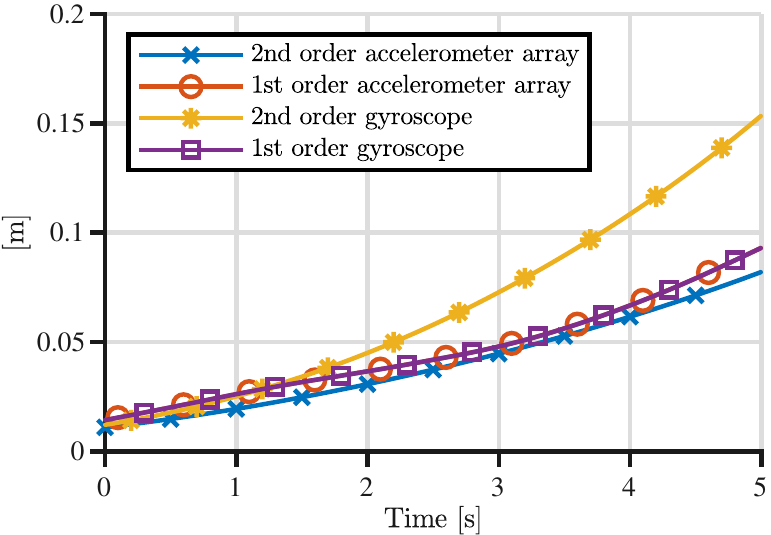}
  }
  \\
  \subfloat[Low dynamics, $F_s = 100$~Hz\label{fig:simulations_pos_rmse_low_dynamics_low_Fs}]{%
    \includegraphics[width=\columnwidth]{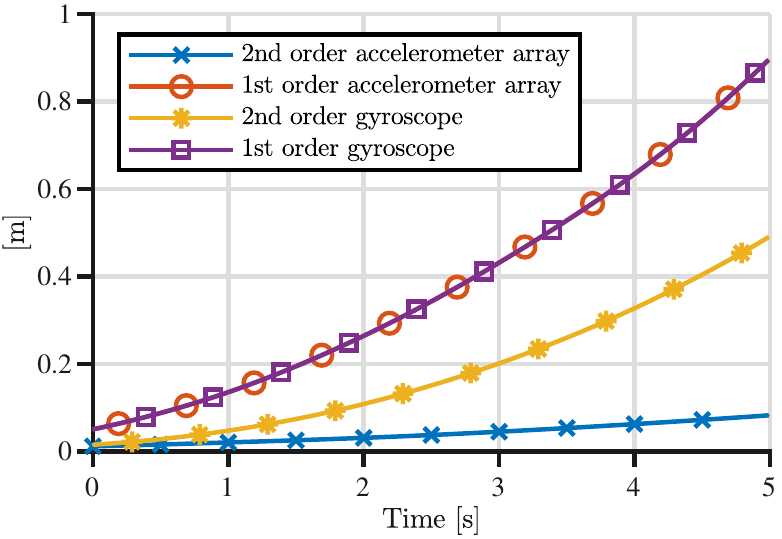}
  }
  \hfil
    \subfloat[High dynamics, $F_s = 100$ Hz\label{fig:simulations_pos_rmse_high_dynamics_low_Fs}]{%
    \includegraphics[width=\columnwidth]{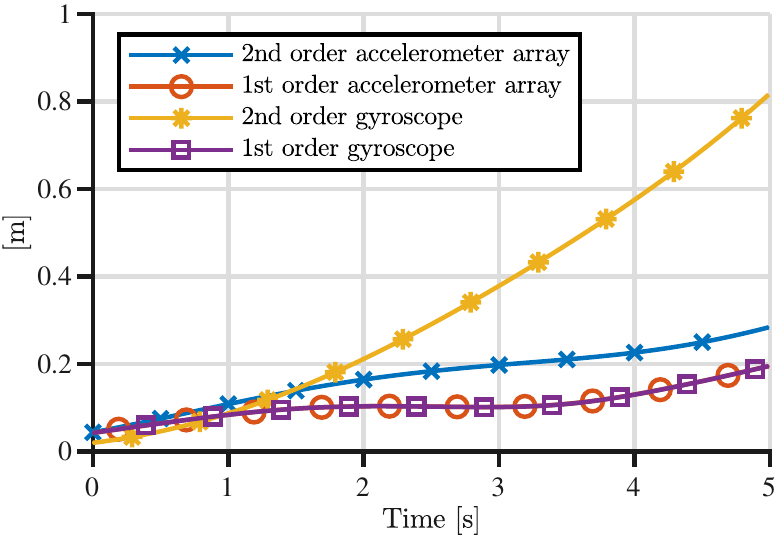}
  }
  \caption{The position estimation RMSE of the four proposed state-spaces
    models in \eqref{eq:2nd_order_accelerometer_array} to
    \eqref{eq:1st_order_gyroscope} using simulated
    data. \protect\subref{fig:simulations_pos_rmse_low_dynamics_high_Fs},
    \protect\subref{fig:simulations_pos_rmse_high_dynamics_high_Fs},
    \protect\subref{fig:simulations_pos_rmse_low_dynamics_low_Fs}, and
    \protect\subref{fig:simulations_pos_rmse_high_dynamics_low_Fs}
    show the RMSE for different rotational dynamics as shown in
    Fig.~\ref{fig:simulations_angular_velocity} and for different
    sampling frequencies $F_s$.}\label{fig:simulations_pos_rmse}
\end{figure*}

We evaluated the performance of the proposed inertial navigation system
realizations in Fig.~\ref{fig:state_space_models} using simulated
data corresponding to different sampling frequencies $F_s$, and
to different degrees of rotational dynamics. The inertial sensor array
considered
in the simulations is shown in Fig.~\ref{fig:array}, which is an
inertial sensor array with 32 \glspl{IMU} attached to a
\gls{PCB}. The accelerometers and
gyroscopes are assumed to be perfectly calibrated in terms of scale
factors, misalignment, cross-coupling, and accelerometer
positions~\cite{Carlsson2021, Carlsson2017, Carlsson2021a}. Hence, the
sensor measurements are
assumed only to be corrupted by a constant bias and white
noise. Specifically, we assume the noise
of the accelerometers and the gyroscopes to be
zero-mean, uncorrelated and to have a standard deviation of $\sigma^{(\text{a})}
= \num{0.5}~\si{m/s^2}$ and $\sigma^{(\text{g})} =
\num{1}~\si{\deg/s}$, respectively. The constant bias is assumed to be
drawn from a normal
distribution with standard deviation set to $\sigma^{(\text{a})}$ and
$\sigma^{(\text{g})}$ for the accelerometers and gyroscope,
respectively. These parameters
were selected to
reflect the performance of typical \gls{MEMS}-based
\glspl{IMU}~\cite{2013-MPU-9150-spec}. Further, we assume idealistically
that the sensors have an infinite dynamic
range.

Field conditions were simulated by generating a continuous
sinusoidal motion from which artificial sensor measurements were
constructed. To account
for the unknown constant bias, artificial position
measurements were
simulated for an initial time period to let the bias estimate of the
filters converge. After
convergence, the position updates ceased and the filters conducted pure
inertial navigation. More specifically, the angular velocity in the
simulations was
set to two different sinusoids as shown in
Fig.~\ref{fig:simulations_angular_velocity},
depicting low dynamic rotational and a high dynamic
rotational motion. From the generated continuous-time motion sequence
discrete time sensor measurements were
generated considering two different sampling frequencies, $F_s = 500~\si{Hz}$ and
$F_s = 100~\si{Hz}$, designating high and low sampling frequencies,
respectively. The synthetic navigation position
measurements were generated from the true motion and with an added
zero-mean noise with a standard
deviation of $\sigma^{(p)} =\num{10}~\si{cm}$ and with a frequency of
$100~\si{Hz}$. The filters had $\sigma^{(b,\text{g})}$,
$\sigma^{(b,s)}$ and $\sigma^{(b,\dot{\omega})}$ all equal to zero. The
initial values for the accelerometer and gyroscope biases were set to zero
and the corresponding initial covariances were set to
$\sigma^{(b,\text{g})}_0 = 3 \sigma^{(\text{g})}$,
$\sigma^{(\text{a})}_0 = 3 \sigma^{(\text{a})}$, with $\sigma^{(b,s)}_0$
and $\sigma^{(b,\dot{\omega})}_0$ evaluated
using~\eqref{eq:covariance_acc_reduced}.

The \gls{RMSE} of the navigation position of the filters during the
inertial navigation phase is shown in
Fig.~\ref{fig:simulations_pos_rmse} for 1000 Monte Carlo
simulations. We first observe that the 1st order state-space
models in~\eqref{eq:1st_order_accelerometer_array}
and~\eqref{eq:1st_order_gyroscope}
give similar results for all considered cases meaning that fusing the
accelerometer array's and the gyroscopes' measurements for angular
velocity estimation yields little improvement over integrating the
angular velocity directly from the gyroscope measurements. Moreover,
for the low dynamic rotational motion, we observe
a reduced error of the 2nd order models in
\eqref{eq:2nd_order_accelerometer_array} and
\eqref{eq:2nd_order_gyroscope} meaning that the 2nd
order numerical integration of the rotation matrix using the angular
acceleration reduces the overall position error growth. However, this
is not the case for highly dynamic rotational motion, where the
reduced error diminishes and even exceeds that of
the 1st order models in~\eqref{eq:1st_order_accelerometer_array}
and~\eqref{eq:1st_order_gyroscope}
for low sampling frequencies. This could be
attributed to the fact that the angular acceleration estimate has its
lowest variance when the angular velocity is zero~\cite{Skog2016}, and
thus the 2nd order integration scheme for the rotation matrix becomes
more accurate for low dynamic rotational motions. With increasing
rotational motion, the variance in angular acceleration increases and
the 2nd order term in the rotation matrix becomes increasingly less
accurate. The same conclusions seem to hold for both high and low
sampling frequencies, indicating that the numerical integration of the
angular velocity is not unstable. Evaluating the poles of the Jacobian
to \eqref{eq:angular_acceleration_body_frame} using the nominal values
of the accelerometer positions of the inertial sensor array yields
three eigenvalues on the imaginary axis, meaning that the
time-continuous equations are stable.

\subsection{Experiments}
\label{sec:experiments}

\begin{figure}[!t]
  \centering
  \subfloat[High dynamics\label{fig:exp_w_motion_high}]{%
    \includegraphics[width=0.97\columnwidth]{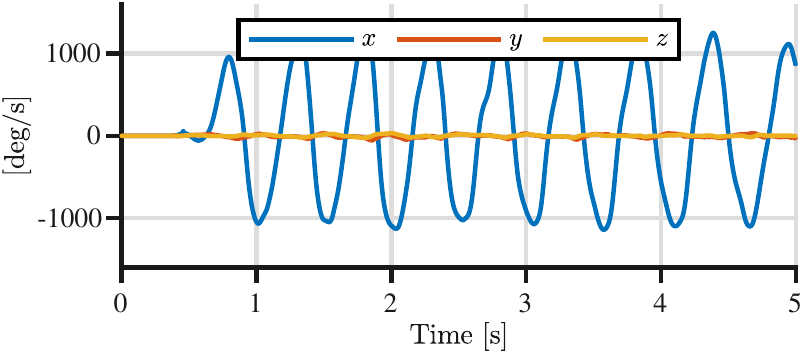}
  }
  \\
  \subfloat[Low dynamics\label{fig:exp_w_motion_low}]{%
    \includegraphics[width=0.97\columnwidth]{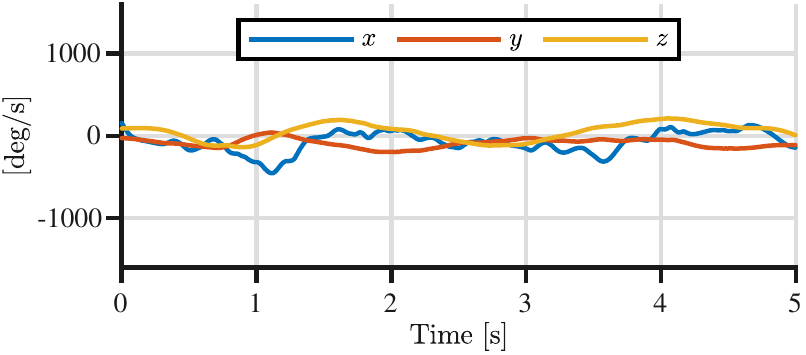}
  }
  \caption{The angular velocity computed by the mean of 32 gyroscope
    triads from experiment 1, where \protect\subref{fig:exp_w_motion_high} and
    \protect\subref{fig:exp_w_motion_low} represent high and low rotational
    dynamics, respectively.}\label{fig:exp_angular_velocity}
\end{figure}

\begin{figure*}[!t]
  \centering
  \subfloat[Low dynamics, $F_s = 500$ Hz\label{fig:experiments_pos_rmse_low_dynamics_high_Fs}]{%
    \includegraphics[width=\columnwidth]{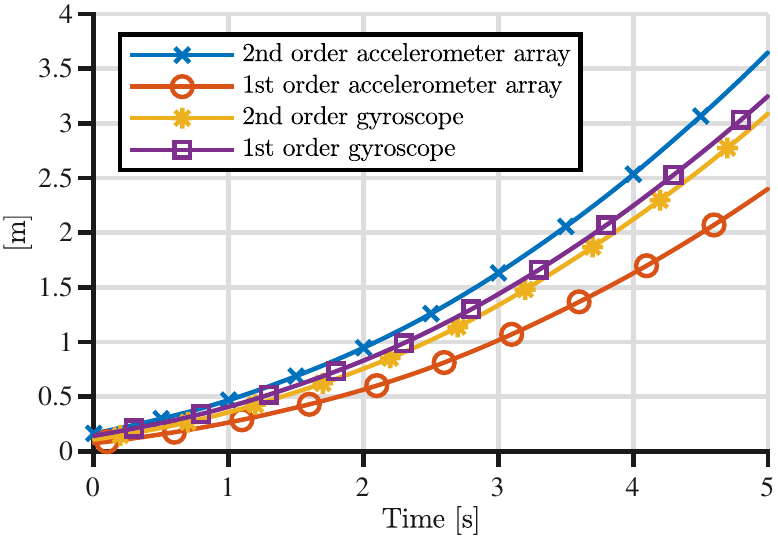}
  }
  \hfil
    \subfloat[High dynamics, $F_s = 500$ Hz\label{fig:experiments_pos_rmse_high_dynamics_high_Fs}]{%
    \includegraphics[width=\columnwidth]{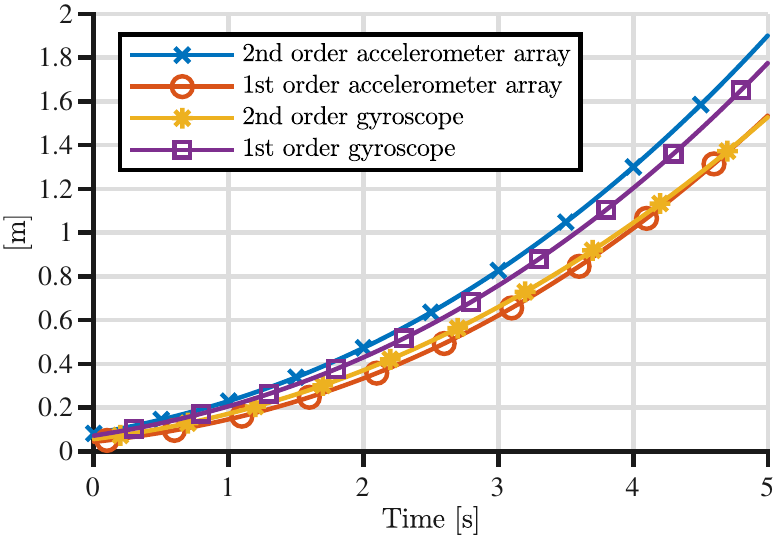}
  }
  \\
  \subfloat[Low dynamics, $F_s = 100$ Hz\label{fig:experiments_pos_rmse_low_dynamics_low_Fs}]{%
    \includegraphics[width=\columnwidth]{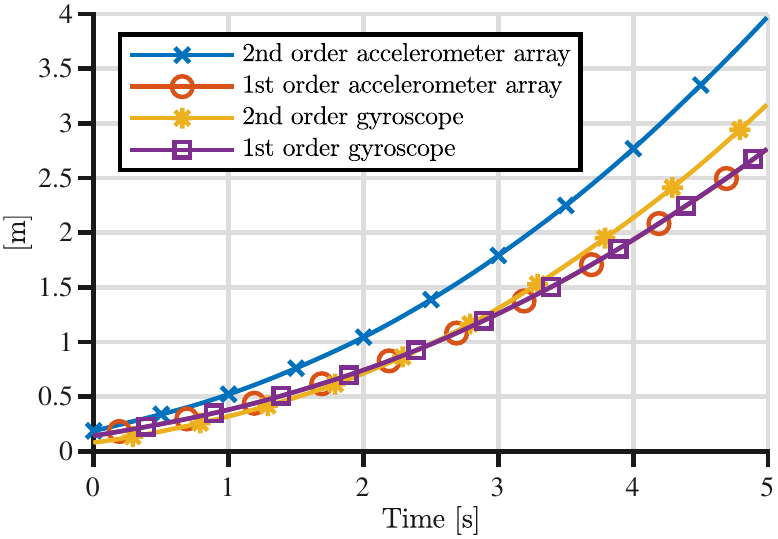}
  }
  \hfil
  \subfloat[High dynamics, $F_s = 100$ Hz\label{fig:experiments_pos_rmse_high_dynamics_low_Fs}]{%
    \includegraphics[width=\columnwidth]{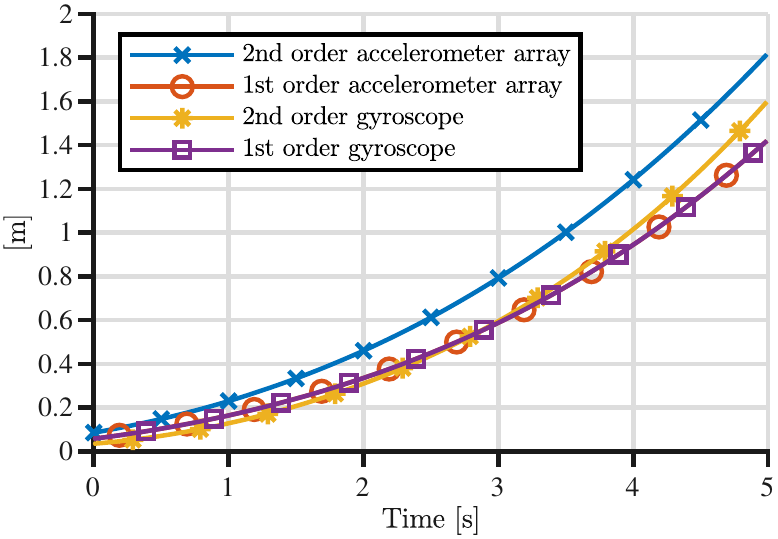}
  }
  \caption{The position estimation RMSE of the four proposed state-spaces
    models in \eqref{eq:2nd_order_accelerometer_array} to
    \eqref{eq:1st_order_gyroscope} using experimental
    data. \protect\subref{fig:experiments_pos_rmse_low_dynamics_high_Fs},
    \protect\subref{fig:experiments_pos_rmse_high_dynamics_high_Fs},
    \protect\subref{fig:experiments_pos_rmse_low_dynamics_low_Fs}, and
    \protect\subref{fig:experiments_pos_rmse_high_dynamics_low_Fs}
    show the RMSE for different rotational dynamics as shown in
    Fig.~\ref{fig:exp_angular_velocity} and for different
    sampling frequencies $F_s$.}\label{fig:experiments_pos_rmse}
\end{figure*}

We also evaluated the performance of proposed inertial navigation system
realizations in Fig.~\ref{fig:state_space_models} using real
experiments performed on the array in Fig.~\ref{fig:array}. We
collected measurements from the array while simultaneously measuring
the position of the array using a camera-based \gls{MC}
system\footnote{\url{https://www.qualisys.com/}}. We assume that
the inertial sensor array samples all the \glspl{IMU} simultaneously
and instantaneously during one time sample. The details about the
hardware and the sampling
process can be found in~\cite{Skog2014b}. The inertial sensor array was
attached to a rig that had multiple reflective markers that were
tracked by the cameras.
The \gls{MC} system provided both a rotation and position estimate of
the rig with
a sampling frequency of $100~\si{Hz}$.

The rig was then exposed to
twisting motion in seven different
experiments. Each experiment consisted of three phases, a
time-synchronization phase, a low rotational dynamics phase, and a high
rotational dynamics phase. The time-synchronization phase
consisted of the rig and the inertial sensor array being rotated around
a single axis. This
rotating motion generated a high \gls{SNR} signal from the gyroscopes, which
was compared with the computed angular velocity from the rotation
estimates given by the \gls{MC} system. An illustration of the rotational
motion during the low and high
dynamic phases is shown in
Fig.~\ref{fig:exp_angular_velocity}, where the angular velocity of the gyroscopes
in experiment 1 is shown. Due to difficulties in aligning the
origin of the \gls{MC} frame and the origin of the inertial
sensor array, the standard deviation for the position updates was set
to $\sigma^{(p)} =\num{10}~\si{cm}$.

In each experiment, the four filters were used to process the measurement from the
inertial sensor array with the same noise variances settings as used in the
simulations in Section~\ref{sec:simulations}. Position measurements
given by the \gls{MC} system were also fed to the filters up to a
certain time point, and then the
position updates ceased. The filters then performed pure inertial navigation
for 5 seconds. The accelerometer and gyroscope biases were initialized
to zero, and
the filters ran for a long enough time period for the bias estimates to
converge before the pure inertial navigation phase was
started. Inertial navigation
commenced from 20 equally spaced time-instants during the low dynamic and
high dynamic phases. For each
rotational phase the \gls{RMSE} during the inertial navigation phase
was computed by averaging the position over the
140 times the position updates ceased and over the coordinate
axes.

The \gls{RMSE} of the position estimates for the experiments is shown in
Fig.~\ref{fig:experiments_pos_rmse} for two different sampling
frequencies. First, we note that
the 2nd order accelerometer array model in
\eqref{eq:2nd_order_accelerometer_array} has a higher \gls{RMSE}
compared to all other models for all considered cases. And the expected
improvement of the 2nd order models in
\eqref{eq:2nd_order_accelerometer_array} and
\eqref{eq:2nd_order_gyroscope} is not seen as observed in the
simulations. Thus, joint propagation of the angular velocity
and the rotation matrix using the angular acceleration
has a higher error than solely propagating the rotation matrix only
using the gyroscope measurements. This
suggests that there is an estimation error in the angular acceleration
computed with the accelerometer array measurements, which could depend on
several factors such as misspecified
covariances, scale factor errors, sensitivity axes misalignments,
cross-coupling terms, and uncertainties in the accelerometer
positions. However,
for the high-frequency sampling case, the 2nd order gyroscope
model in \eqref{eq:2nd_order_gyroscope} and the 1st order accelerometer
array in \eqref{eq:1st_order_accelerometer_array} have a
lower error compared to the 1st order
gyroscope model in \eqref{eq:1st_order_gyroscope}, meaning that the
estimated angular acceleration
can improve the orientation estimation. And for the 100 Hz
sampling frequency, we observe that the \gls{RMSE} for the accelerometer array
models in \eqref{eq:2nd_order_accelerometer_array} and
\eqref{eq:1st_order_accelerometer_array} increase relative to the 1st
order gyroscope
model in~\eqref{eq:1st_order_gyroscope}. Thus, when the
integration time increases the error in the orientation estimation
using the angular
acceleration increases, suggesting that the angular velocity
integration is not stable. Since the simulations did not suggest
an unstable integration for the considered geometry of the accelerometer
positions, it is left for future research to find the cause for the
observed behavior.

\section{Summary and Conclusions}
\label{sec:conclusion}

We propose four state-space models of how to perform
inertial navigation using an inertial sensor array. Using different
underlying assumptions regarding the orientation modeling, these four
models fuse the measurements of multiple and
generally placed accelerometer triads
and a gyroscope triad to improve inertial navigation
performance. The non linear problem of fusing the
accelerometer array measurements and the
gyroscope measurements for angular velocity estimation is solved
through Kalman updates. The benefit of using Kalman updates is
twofold. First, the required computations for the sensor fusion
problem can be known in advance, which is not the case when this
fusion is done using an
iterative optimization method~\cite{Skog2016}. Knowing the number of
computations in advance is useful
for embedded systems where computational resources are
limited. Second, we show that the Kalman updates stabilize
the potentially unstable equations when the angular velocity is
integrated using the accelerometer array
measurements~\cite{Nusbaum2017,Padgaonkar1975}. Thus, it is not
necessary to place the accelerometer triads in complex geometries to
ensure stability~\cite{Park2005,
  Hanson2005,Sahu2020,Williams2013,Chen1994a}. Furthermore, the
angular acceleration is estimated from the
accelerometer array measurements
and we present three methods for how to capitalize on the angular
acceleration: \begin{inparaenum}[1)]
either \item  propagate the angular velocity; or
            \item propagate the rotation matrix with 2nd order
              accuracy; or
            \item both.
        \end{inparaenum} We also show that
the correct weighting of multiple accelerometer measurements for
computation of the specific force is given by the mean
value~\cite{Skog2014}, which is
derived using the implicit assumption of centered accelerometer
positions. Moreover, we show how only six dimensions of all individual
accelerometer biases are observable and
consequently, it is sufficient to only model the bias in six dimensions
independent of the number of accelerometer triads. This
observation answers the question posed
in~\cite{Waegli2008} where it was observed that the individual biases
of the accelerometers were unobservable, and that only three
dimensions of all bias terms were observable when computing the
specific force.

Moreover, simulations showed that for low rotational motions a
significant reduction in
navigation position error can be achieved compared to the conventional
inertial navigation method when
using all available information from the
accelerometer array and the gyroscopes. This
can be attributed to the increased accuracy in the
numerical integration of the rotation matrix by the inclusion of a 2nd
order term given by the angular acceleration.
However, this error reduction was not observed
in real-world experiments. This can be ascribed to errors
in the assumed covariances, accelerometer positions, scale-factors,
and misalignments. Thus, future research should focus on methods to calibrate
and
compensate for these error sources. However, it was observed from the
experiments that the inertial sensor array could, for high sampling
frequencies, provide extra orientation information and reduce the
inertial navigation error compared to conventional inertial navigation.

\appendices

\section{Accelerometer Array Differential Equations}
\label{sec:array-equations}

Concatenating the measurements from $K$ accelerometer triads
using~\eqref{eq:pointwise_acceleration_triad} yields
\begin{equation}
  f^b = h(\omega^b, r_{1:K}^{b}) + H(r_{1:K}^{b})  \begin{bmatrix}
    \dot{\omega}^b \\
    s^b
  \end{bmatrix}
\end{equation}
where $f^b$ and $h$ are defined in~\eqref{eq:concat_specific_force}
and
\begin{equation}
  H(r_{1:K}^{b})
    \triangleq \begin{bmatrix}
    -[ r_1^b \times ] & I_3 \\
    \vdots & \vdots \\
    -[r_K^b \times ] & I_3
  \end{bmatrix}.
\end{equation}
The solution for the angular accelerations and the specific force is then
\begin{equation}
  \label{eq:8}
  \begin{bmatrix}
    \dot{\omega}^b \\
    s^b
  \end{bmatrix}
  =
  A \left(f^b - h(\omega^b, r_{1:K}^{b})  \right),
\end{equation}
where
\begin{equation}
  \label{eq:13}
  A \triangleq \left( H^\top H \right)^{-1} H^\top.
\end{equation}
The angular acceleration and the specific force are thus
different linear combinations determined by the matrix $A$ of the difference between the
specific forces $f_k$ and the centrifugal accelerations $h(\omega^b,
r_{1:K}^{b})$. Equation~\eqref{eq:8} can be simplified by noting that
\begin{equation}
  \label{eq:10}
  H^\top H = \begin{bmatrix}
    \sum_k [ r_k^b \times ]^\top [ r_k^b \times ] & - \sum_k [ r_k^b
    \times ]^\top \\
    -\sum_k [ r_k^b \times ] & K I_3
  \end{bmatrix}.
\end{equation}
Assuming that $\sum_k  r_k^b = 0$, and consequently $\sum_k [ r_k^b
\times ] = 0$, yields
\begin{equation}
  \label{eq:11}
  H^\top H = \begin{bmatrix}
    \sum_k [ r_k^b \times ]^\top [ r_k^b \times ] & 0
     \\
    0  & K I_3
  \end{bmatrix}.
\end{equation}
The three last rows of \eqref{eq:8} then become
\begin{equation}
    \frac{1}{K}\sum_{k = 1}^K ( f_k^b - [\omega^b \times ]^2 r_k^b ) = \frac{1}{K}\sum_{k = 1}^K  f_k^b,
\end{equation}
which are independent of $\omega^b$. The differential equations
in~\eqref{eq:8} then decouples and become~\eqref{eq:diff_eq_continuous_array_ext}.

\section{Time Evolution of Rotation Vector}
\label{sec:time-evol-rotat}

To show that \eqref{eq:lie_algebra_diff_eq_solution} is equal to
\eqref{eq:9} amounts to calculating the terms $\dot{\theta}(0)$ and
$\ddot{\theta}(0)$. The rotation vector can be expressed as $\theta =
\|\theta\| u$ where $\|\theta\|$ is the magnitude of $\theta$ and $u$ is the unit
vector pointing in the same direction as $\theta$. Using this,~\eqref{eq:27} can be
rewritten as
\begin{equation}
  \label{eq:2}
  \Gamma(\theta) = I_3 + \frac{1}{2}\|\theta\|[u\times] + \left(1 -
    \frac{\|\theta \|}{2} \cot \left(
      \frac{\| \theta \|}{2}\right) \right)[u \times]^2.
\end{equation}
It is then clear that $\Gamma(0) = I_3$. With $\theta(0) = 0$
the first time-derivative of $\theta$ is
\begin{equation}
  \label{eq:1}
  \dot{\theta}(0) = \Gamma(\theta(0))\omega(t) = \omega(t).
\end{equation}
Next $\ddot{\theta}(0)$ is obtained through
\begin{equation}
  \label{eq:3}
  \ddot{\theta}(\tau) = \dot{\Gamma}(\theta(\tau))\omega(t + \tau) + \Gamma(\theta(\tau))\dot{\omega}(t + \tau).
\end{equation}
By differentiation, it can be shown that
\begin{equation}
  \label{eq:4}
  \left.\dot{\Gamma}(\theta(\tau))\right|_{\tau = 0} = -\frac{1}{2}[\omega(t)\times],
\end{equation}
so that
\begin{equation}
  \label{eq:5}
  \ddot{\theta}(0) = -\frac{1}{2}[\omega(t)\times] \omega(t) +
  \dot{\omega}(t) = \dot{\omega}(t).
\end{equation}

\section{Jacobians}
\label{sec:jacobians}
The Jacobian in~\eqref{eq:D-LG-EKG-propagation-jacobians-state}
to~\eqref{eq:20} in the state-matrix $X_n$ is given by
\begin{equation}
  \label{eq:12}
  J_n^{(x)}
  = \begin{bmatrix}
    J_n^{(1)}
    &
    J_n^{(2)}
    \\
    0_{9,12}
    &
    0_{9}
  \end{bmatrix},
\end{equation}
where
\begin{subequations}
\begin{align}
  \label{eq:25}
  J_n^{(1)}
  &
    = \begin{bmatrix}
    0_3 
    &
    I_3T + \frac{\partial \dot{\omega}_{n} }{\partial
      \omega} \frac{T^2}{2}
    &
    0_3
    &
    0_3
    \\ 
    0_3
    &
    \frac{\partial \dot{\omega}_{n} }{\partial \omega} T
    &
    0_3
    &
    0_3
    \\ 
    \frac{\partial \dot{v}_{n} }{\partial
      e_{R}} \frac{T^2}{2}
    &
    0_3
    &
    0_3
    &
    I_3 T
    \\ 
    \frac{\partial \dot{v}_{n} }{\partial e_{R}} T
    &
    0_3
    &
    0_3
    &
    0_3
  \end{bmatrix},
\end{align}
and
\begin{align}
  J_n^{(2)}
  &
    = \begin{bmatrix}
      \frac{\partial \dot{\omega}_{n} }{\partial
        b^{(\dot{\omega})}} \frac{T^2}{2}
    &
    0_{3}
    &
    0_{3}
    \\ 
    \frac{\partial \dot{\omega}_{n} }{\partial
      b^{(\dot{\omega})}} T
    &
    0_{3}
    &
    0_{3}
    \\ 
    0_{3}
    &
    \frac{\partial \dot{v}_{n} }{\partial
      b^{(s)}} \frac{T^2}{2}
    &
    0_{3}
    \\ 
    0_{3}
    &
    \frac{\partial \dot{v}_{n} }{\partial
      b^{(s)}} T
    &
    0_{3}
  \end{bmatrix}.
\end{align}
\end{subequations}
Here $e_{R} \in \R^3$ is the Lie algebra variable to $R$. The partial
derivatives of $\dot{\omega}_{n}$ are given by
\begin{subequations}
\begin{align}
  \frac{
  \partial \dot{\omega}_{n}
  }{
  \partial
  \omega
  }
  &
    = \sum_kA_k\left( [ [\omega_n \times ]r_k \times ] + [\omega_n
    \times][r_k \times]  \right),
  \\
  \frac{\partial \dot{\omega}_{n} }{\partial
  b^{(\dot{\omega})}}
  &
    = I_3,
\end{align}
\end{subequations}
derivatives of $\dot{v}_{n}$ are given by
\begin{align}
  \frac{\partial \dot{v}_{n} }{\partial
  e_{R}}
  &
    =
    -R_n \left[ s_n  \times \right],
  &
  \frac{\partial \dot{v}_{n} }{\partial
  b^{(s)}}
  &
    = I_3.
\end{align}
The Jacobian
in~\eqref{eq:D-LG-EKG-propagation-jacobians-noise} to~\eqref{eq:20} in
the noise vector $w_n^{(p)}$ is given by
\begin{subequations}
\begin{align}
J_n^{(w)}
  &
    =
    \begin{bmatrix}
      \frac{
        \partial \dot{\omega}_{n}
      }{
        \partial w_{n}^{(\dot{\omega})}
      } \frac{T^2}{2}
      &
      0_{3,9}
      \\
      \frac{\partial \dot{\omega}_{n} }{\partial
        w_{n}^{(\dot{\omega})}} T
      &
      0_{3, 9}
      \\
      \frac{\partial \dot{v}_{n} }{\partial
        w_{n}^{(s)}} \frac{T^2}{2}
      &
      0_{3,9}
      \\
      \frac{\partial \dot{v}_{n} }{\partial
        w_{n}^{(s)}} T
      &
      0_{3,9}
      \\
      0_{9, 3} & I_{9}
    \end{bmatrix}
\end{align}
\end{subequations}
where the partial derivatives are given by
\begin{align}
  \frac{\partial \dot{\omega}_{n} }{\partial
  w_{n}^{(\dot{\omega})}}
  &
    = I_3,
  &
  \frac{\partial \dot{v}_{n} }{\partial
  w_{n}^{(\text{a})}}
  &
    = I_3.
\end{align}


\ifCLASSOPTIONcaptionsoff
  \newpage
\fi



\bibliographystyle{IEEEtran}
\bibliography{ref}





%




\end{document}